\documentclass{aa}
\usepackage[varg]{txfonts}
\pdfoutput=1
\usepackage{graphicx}
\usepackage{newtxtext,newtxmath}
\usepackage{xcolor}
\usepackage{verbatim}
\usepackage{subcaption}
\usepackage{ulem}
\RequirePackage[colorlinks=true,linkcolor=blue,citecolor=blue,urlcolor=blue]{hyperref}

\graphicspath{{figures/}}

\newcommand{\mvir}{M_{\rm 200c}}
\newcommand{\mvirhost}{\mvir^{\rm host}}
\newcommand{\rvir}{R_{\rm 200c}}
\newcommand{\rvirhost}{\rvir^{\rm host}}
\newcommand{\mgas}{M_{\rm gas}}
\newcommand{\mgassat}{\mgas^{\rm sat}}
\newcommand{\mcgas}{M_{\rm ColdGas}}
\newcommand{\mcgassat}{\mcgas^{\rm sat}}
\newcommand{\mhgas}{M_{\rm HotGas}}
\newcommand{\mhgassat}{\mhgas^{\rm sat}}

\newcommand{\mstar}{M_\star}
\newcommand{\mstarsat}{\mstar^{\rm sat}}
\newcommand{\mcgm}{M_{\rm CGM}}
\newcommand{\mcgmsat}{\mcgm^{\rm sat}}

\newcommand{\rhalfstar}{R_{\rm half,\star}}

\newcommand{\dsathost}{d_{\rm sat}^{\rm host}}
\newcommand{\sublink}{\textsc{sublink}}
\newcommand{\sublinkgal}{\sublink\_\textsc{gal} }
\newcommand{\subfind}{\textsc{subfind} }

\newcommand{\msun}{M_\odot}

\makeatletter
\renewcommand*\aa@pageof{Page \thepage{} of \pageref*{LastPage}}
\makeatother

\begin{document}

\title{The hot circumgalactic media of massive cluster satellites in the TNG-Cluster simulation: Existence and detectability}

\titlerunning{The CGM of Massive Satellites in TNG-Cluster}

\author{Eric Rohr\inst{1}\thanks{Contact e-mail: \href{mailto:rohr@mpia.de}{rohr@mpia.de}}\orcid{0000-0002-9183-5593}
\and Annalisa Pillepich\inst{1}\orcid{0000-0003-1065-9274}
\and Dylan Nelson\inst{2}\orcid{0000-0001-8421-5890}
\and Mohammadreza Ayromlou$^2$\orcid{0000-0003-3783-2321}
\and Elad Zinger$^{3,1}$\orcid{0000-0002-6316-3996}
}

\institute{Max-Planck-Institut f{\"u}r Astronomie, K{\"o}nigstuhl 17, D-69117 Heidelberg, Germany \label{1}
\and Zentrum f{\"u}r Astronomie der Universit{\"a}t Heidelberg, ITA, Albert Ueberle Str. 2, D-69120 Heidelberg, Germany \label{2}
\and Centre for Astrophysics and Planetary Science, Racah Institute of Physics, The Hebrew University, Jerusalem 91904, Israel \label{3}
}

\date{}

\abstract{
The most massive galaxy clusters in the Universe host tens to hundreds of massive satellite galaxies~$\mstar\sim10^{10-12.5}\,\msun$, but it is unclear if these satellites are able to retain their own gaseous atmospheres. We analyze the evolution of $\approx90,000$ satellites of stellar mass $\sim10^{9-12.5}\,\msun$ around 352 galaxy clusters of mass $\mvir\sim10^{14.3-15.4}\,\msun$ at $z=0$ from the new TNG-Cluster suite of cosmological magneto-hydrodynamical galaxy cluster simulations. The number of massive satellites per host increases with host mass, and the mass--richness relation broadly agrees with observations. A halo of mass $\mvirhost\sim10^{14.5}\,(10^{15})\,\msun$ hosts $\sim100\,(300)$ satellites today. Only a minority of satellites retain some gas, hot or cold, and this fraction increases with stellar mass. lower-mass satellites $\sim10^{9-10}\,\msun$ are more likely to retain part of their cold interstellar medium, consistent with ram pressure preferentially removing hot extended gas first. At higher stellar masses $\sim10^{10.5-12.5}\,\msun$, the fraction of gas-rich satellites increases to unity, and nearly all satellites retain a sizeable portion of their hot, spatially extended circumgalactic medium~(CGM), despite the ejective activity of their supermassive black holes. According to TNG-Cluster, the CGM of these gaseous satellites can be seen in soft X-ray emission~(0.5-2.0~keV)~that is,$\gtrsim10$~times brighter than the local background. This X-ray surface brightness excess around satellites extends to $\approx30-100$~kpc, and is strongest for galaxies with higher stellar masses and larger host-centric distances. Approximately $10$~percent of the soft X-ray emission in cluster outskirts $\approx0.75-1.5\rvir$ originates from satellites. The CGM of member galaxies reflects the dynamics of cluster-satellite interactions and contributes to the observationally inferred properties of the intracluster medium.
}


\maketitle

\section{Introduction} \label{sec:intro}

Since the 1950s, astronomers have been observing absorption lines in the spectra of background sources due to gas along the line of sight, including in the gaseous halos of intervening galaxies. This circumgalactic medium (CGM) is an enigmatic component of the baryonic Universe, and a necessary element for a comprehensive picture of galaxy evolution \citep[see][for recent reviews]{Tumlinson2017,Faucher-Giguere2023}. The CGM is the interface between gas that is expelled from the galaxy due to stellar and active galactic nucleus (AGN) feedback, and gas flowing into the halo from the intergalactic medium (IGM), as well as from satellite galaxies (hereafter satellites). 

To first approximation, the CGM is a thermally pressure-supported atmosphere in rough hydrostatic equilibrium \citep{Rees1977,Silk1977,White1978}. With increasing halo mass and virial temperature, the CGM also becomes hotter and emits at progressively higher energies. Inflowing gas is expected to be shock-heated to approximately the same virial temperature or, depending on galaxy mass and redshift, to penetrate toward the inner regions \citep[][]{Keres2005,Dekel2006,Faucher-Giguere2011,Nelson2016}. However, outflows, inflows, and the quasi-static CGM are all multiphase, and can contain cold, molecular components as well as hot, virialized phases. As a result, the CGM is a complex component of the galactic ecosystem that is shaped by the complex interplay of physical processes including gas cooling, magnetic fields, galactic feedback, filamentary inflows, and galaxy mergers and satellites \citep[e.g.,][]{Sarazin2002,Voit2005,Peeples2019,Nelson2020,Ramesh2023a}. Within the current cosmological paradigm, all massive central galaxies (hereafter centrals) are expected to be surrounded by a gaseous halo or CGM, embedded within a halo of cold dark matter. The picture for satellites orbiting within the halo of a more massive central galaxy is, however, less clear. It is unknown if satellites can retain their own CGM after they begin interacting with their host halo.

Satellite and central galaxies differ in many respects. Observationally, satellites have higher elliptical and S0 fractions, higher quenched fractions, lower (specific) star formation rates (SFRs), redder colors, lower neutral and molecular gas fractions, elevated gas metallicities, reduced X-ray emission, and suppressed AGN activity compared to centrals of the same mass \citep[e.g.,][]{Dressler1980,Giovanelli1985,Wetzel2013,Peng2010,Peng2012,Brown2016,Maier2019,Maier2019b,Cortese2021,Boselli2022}. A number of physical mechanisms have been invoked to explain these differences \citep[see][for a recent review]{Cortese2021}, many of which are related to the hydrodynamical interaction between the gas of the satellites and the ambient gaseous halo of their hosts. 

Starting from the outermost scales and working inward, satellites are thought to be cut off from the intergalactic medium (IGM), removing a source of gas replenishment \citep[][]{Larson1980}. Satellite gas is vulnerable to ram pressure stripping (RPS), which directly removes gas preferentially from the outside in \citep[][for a recent review]{Gunn1972,McCarthy2008,Boselli2022}. First, the gaseous halo of the satellite is stripped, after which it is no longer able to accrete from its own CGM \citep[][]{Balogh2000}. Moreover, gas accretion from the ambient medium is expected to be strongly suppressed \citep{vandeVoort2017,Wright2020}. Finally, ram pressure can also directly remove the interstellar medium (ISM) of satellites, likely in an outside in fashion as inferred by observations of truncated disks \citep[e.g.,][]{Warmels1988,Cayatte1990,Cayatte1994,Vollmer2001,Lee2022}. Consequently, RPS deposits the CGM and ISM of satellites into the gaseous halos of their hosts \citep[e.g.,][]{Rohr2023,Roy2024}. 

Simultaneously with these environmental effects, the ISM and CGM gas of satellites is subject to internal stellar and AGN feedback processes. These feedback processes increase the effectiveness of RPS \citep[e.g.,][]{Bahe2015,Ayromlou2021b,Kulier2023}. Several observations suggest that ram pressure may compress the ISM and temporarily enhance star formation \citep[e.g.,][]{Gavazzi2001,Vulcani2018,Roberts2022} and AGN activity \citep[e.g.,][contra: \citealt{Roman-Oliveira2019}]{Poggianti2017,Peluso2022} before eventually removing most satellite gas. This ram pressure driven increase in ISM gas density has also been seen in some idealized simulations of individual satellites \citep[][]{Lee2020,Choi2022,Zhu2024} and to cause bursts of star formation also in cosmological volume galaxy simulations \citep{Goeller2023}.

The effectiveness of RPS increases with host mass. Furthermore, at fixed host mass, ram pressure stripping is expected to become more important with decreasing satellite stellar mass, since the satellite stellar body acts as the gravitationally restoring force opposing ram pressure \citep[e.g.,][]{Wright2022,Zinger2024}. However at galaxy stellar masses $\gtrsim10^{10.5}\msun$ AGN feedback is expected and inferred to become more efficient at ejecting gas and quenching galaxies \citep[e.g.,][]{Silk1998,Schawinksi2007,Fabian2011}, both in centrals and satellites \citep[e.g.,][]{Bluck2020,Joshi2020,Donnari2021b}. After the ISM has been largely ejected, it becomes more susceptible to ram pressure removal. As a result, the retention of hot CGM and cold ISM gas reservoirs, on timescales of Myr to Gyr, depends sensitively on both satellite mass as well as host halo mass.

The low density and surface brightness of the CGM have made it difficult to directly observe in emission, but recent instrumentation and surveys have detected the CGM of single and stacked galaxies in the optical/UV \citep[e.g.,][]{Hayes2013,Leclercq2022,Dutta2023}, X-ray \citep[e.g.,][]{Bogdan2013,Comparat2022,Chadayammuri2022}, and even the radio \citep{Chen2024}. For satellites, the first detection of a gaseous corona in emission was the Large Magellanic Cloud (LMC) using {\it Hubble Space Telescope} spectra \citep{Wakker1998,Krishnarao2022}. Using {\it Chandra}, \citet{Sun2007} detect X-ray galactic coronae with temperatures $kT\approx 0.5-1.1$~keV in $\approx60$~percent of super-$L_*$ galaxies residing in clusters, implying CGM stripping timescales of several Gyrs. \citet{Goulding2016} detect X-ray emission within the stellar effective radius of 33 early-type systems in the MASSIVE survey, a fraction of which are not the central of their group or cluster. \citet{Babyk2018} extract X-ray scaling relations out to five times the stellar effective radius of 94 early-type galaxies in the local Universe, many of which are satellites of nearby groups and clusters. \citet{Zhang2019} find in stacked SDSS spectra suppressed H$\alpha$ + [N\, {\sc ii}] emission from the CGM of $\sim{\rm L}_\star$ galaxies that are in denser environments (see also \citealt{Burchett2018}). Lastly, \citet{Hou2024} and \citet{Zhang2024} find X-ray emission around satellites in stacked {\it Chandra} and eROSITA observations (see also \citealt{Hou2021}). However, it remains unclear what percentage of satellites today retain hot gas reservoirs, and to what extent their CGM have been damaged by the environment of their hosts, especially if these are massive groups and clusters of galaxies. 

Despite the relatively low number statistics for detected CGM gas around satellites, its existence is consequential. For example, \citet{Lucchini2020} find that such a satellite CGM is necessary to reproduce the observed kinematics and mass of the Large and Small Magellanic Cloud system. Moreover, the high surviving H{\sc i} fractions of a Hydra galaxy cluster subgroup suggest a surviving intragroup medium actually shields the group members from the hotter intracluster medium \citep[ICM;][]{Hess2022}. Lastly, \citet{Churazov2012} infer that the largest fluctuations in resolved X-ray maps of the Coma cluster are due to cluster members. 

The stripping and loss of gas from satellites have been studied with cosmological hydrodynamical galaxy simulations. For example, using the {\sc gimic} simulations, \citet{Bahe2013} find that group and cluster hosts with total mass $\sim10^{13-15.2}\, \msun$ are able to strip infalling galaxies of their CGM already when the latter are at distances of $\sim5\rvirhost$. More recently, \citet{Wright2022} study the orbital histories of {\sc eagle} satellites around groups and clusters, finding that satellites begin to lose their CGM at $\approx2-3\rvirhost$, while gas removal is more efficient for clusters and lower-mass satellites. The emerging phenomenology of diverse satellite CGM properties and removal timescales depends on satellite-host configurations \citep[e.g.,][]{Kawata2008,McCarthy2008,Bekki2009,Zinger2018,Kulier2023}. This has also led to the revision of simplified assumptions previously adopted in semi-analytical models. Typically, these models have only accounted for gas stripping in satellites within the virial radius \citep[e.g.,][]{Henriques2015}, with some even removing the entire satellite CGM gas once a satellite crosses into the halo boundary \citep[e.g.,][]{Lacey2016,Lagos2018}. Updated semi-analytical models, which include both the stripping of satellites and centrals beyond the halo boundary, as well as a gradual approach to gas stripping, demonstrate improved alignment with observational data \citep[e.g.,][]{Ayromlou2021a}. However, the majority of the theoretical analyses so far focus on galaxy groups and low-mass clusters rather than the largest clusters in the Universe, where environmental effects are maximal. Moreover, no previous simulation work has connected the possible survival of the CGM around satellites and its observability.

In this work, we use the new TNG-Cluster simulation suite \citep[][\textcolor{blue}{Pillepich et al. in prep.}]{Nelson2024} to address the question: do $z=0$ satellites in massive galaxy clusters retain, or not, their CGM. We specifically target the massive end of the host distribution to focus on the harshest environments and to maximize the mass range of satellite galaxies. TNG-Cluster includes 352 galaxy clusters with total mass $\mvirhost\sim10^{14.3-15.4}\,\msun$ and over 90,000 satellites with stellar mass larger than $\sim10^{9}\,\msun$ at $z=0$. It therefore provides an unprecedented sample size of satellites, including thousands of satellites more massive than our own Milky Way and Andromeda. These are simulated with the well-tested IllustrisTNG galaxy formation model (TNG hereafter) within a full $\Lambda$CDM cosmological context, and with competitive spatial and mass resolution. Despite important simplifications (such as no explicit modeling of the multiphase ISM, influencing results related to the cold gas -- see \citealt{Zinger2018} and \citealt{Kukstas2022}), the TNG model returns satellite quenched fractions and gas contents that are broadly consistent with observations \citep[e.g.,][]{Stevens2019,Donnari2021b,Stevens2021}.

Moreover, as shown in a series of companion papers that showcase first results from TNG-Cluster, the simulated halos exhibit X-ray luminosity scaling relations and other global properties of the ICM \citep{Nelson2024}, fractions of cool cores \citep{Lehle2024}, gas kinematics \citep{Ayromlou2024}, levels of turbulence in the cores \citep{Truong2024}, and morphologically diverse radio relics \citep{Lee2024}  that are all broadly consistent with observations.

The goals of this work are: a) to quantify the predictions from TNG-Cluster for the population of cluster satellites and their hot and cold gas reservoirs; by doing so, we aim to b) provide interpretation for current X-ray observations of cluster galaxies and their spatial extent; c) suggest an experiment to constrain the physics of satellites in massive hosts, beyond X-ray observations of individual systems; and d) quantify the covering fraction of the CGM of satellites versus the ICM of the host.

We begin by describing the new simulation suite, sample selection, and our methods (\S~\ref{sec:meth}). In \S~\ref{sec:results}, we present the clusters and their satellite demographics, the gaseous content of the satellites, the spatial extent of the satellite CGM, and the causes of satellite-to-satellite variations. We discuss details and caveats of our results in \S~\ref{sec:discussion}, and in \S~\ref{sec:discussion_xray} we present a statistical stacking experiment to detect the soft X-ray emission from satellite CGM. We discuss implications of our findings in \S~\ref{sec:discussion_absorption} and \S~\ref{sec:discussion_evolution} and summarize our main results in \S~\ref{sec:sum}.

\renewcommand{\arraystretch}{1.5}
\begin{table*}
    \centering
    \caption{
    Summary of the definitions and notation used throughout this work.
    }
    \label{tab:definitions}
    \begin{tabular}{lll} 
        \hline
        Name & Notation & Notes \\ 
        \hline
        host halo size & $\rvirhost$~[Mpc] & \parbox{11cm}{radius at which the average enclosed density is 200 times the critical density of the Universe}\\
        host halo mass & $\mvirhost\, [\msun]$ & total mass enclosed within the halo size $\rvirhost$ \\
	\hline
        satellite stellar size & $\rhalfstar$~[kpc] & \parbox{11cm}{half mass radius of the total gravitationally bound stellar mass, computed via \subfind} \\ 
        satellite stellar mass & $\mstarsat\, [\msun]$ & stellar mass of the satellite galaxy within $2\rhalfstar$ \\
        satellite gas mass & $\mgassat\, [\msun]$ & total gravitationally bound gas mass, computed via \subfind \\
        satellite hot gas mass & $\mhgassat\ \, [\msun] $ & $\mgassat$ but only for hot $>10^{4.5}$~K gas; used as a proxy for X-ray bright gas \\
        satellite cold gas mass & $\mcgassat\ \, [\msun] $ & \parbox{11cm}{$\mgassat$ but only for cold $<10^{4.5}$~K gas; used as a proxy for ISM gas} \\
        satellite CGM gas mass & $\mcgmsat\ \, [\msun] $ & \parbox{11cm}{circumgalactic medium; $\mgassat$ but only for $>2\rhalfstar$ gas; includes hot and cold gas} \\ 
        \hline
        host-centric distance & $\dsathost\, [\rvirhost,\rm{kpc}]$ & \parbox{11cm}{distance from the center of the host cluster to the center of the satellite, in 2D projection unless otherwise noted.} \\ 
        \hline
        infall time & $\tau_{\rm FirstInfall}$~[Gyr] & the first time a galaxy became a satellite$^a$ \\
    \end{tabular}
    \parbox{\textwidth}{
        \footnotesize{
        $^a$The first time the galaxy was a Friends-of-Friends satellite for at least 3 consecutive snapshots in a host of mass $\mvirhost > 10^{12}\, \msun$. \\ 
        }
    }
\end{table*}

\begin{figure*}
    \includegraphics[]{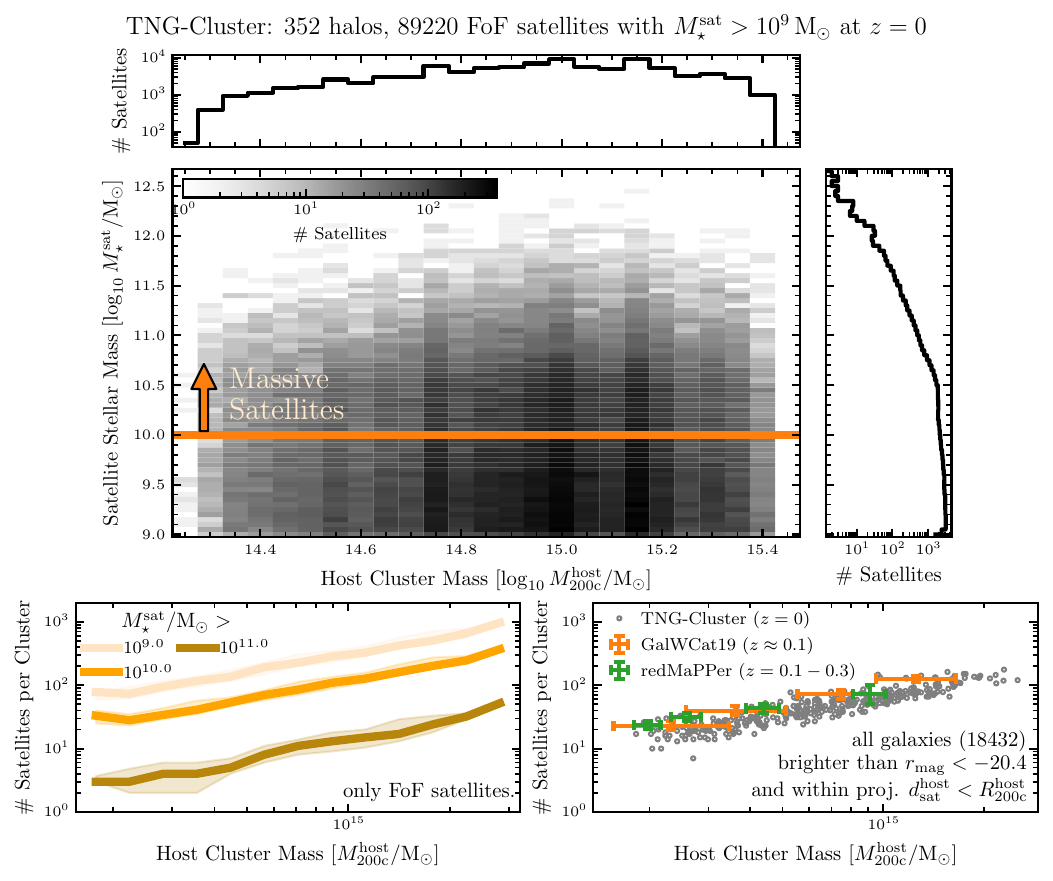}
    \caption{
    {\bf Cluster and satellite demographics in TNG-Cluster.}
    {\it Top panels:} We plot the number of satellite galaxies of the 352  clusters in bins of satellite stellar $\mstarsat$ and host halo $\mvirhost$ mass at $z=0$. The one-dimensional histograms for $\mstarsat$ and $\mvirhost$ are included in the right and top panels. The number of satellites per stellar mass bin decreases with satellite stellar mass, where there is a sharp decrease in the number of super-L$_\star$ galaxies $\mstarsat>10^{10.5}\, \msun$. The number of satellites per host mass bin increases at lower cluster masses $\mvirhost\sim10^{14.3-15}\,\msun$, and then decreases for the most massive clusters $\mvirhost\sim10^{15-15.4}\,\msun$, which reflects the TNG-Cluster halo mass function \citep[see text and][for details]{Nelson2024}.
    {\it Bottom panels:} For each of the 352 hosts, we plot the number of FoF satellites above given stellar masses (left), where the medians and 16th and 84th percentiles are the solid curves and shaded regions respectively. Additionally, we qualitatively compare the TNG-Cluster results to spectroscopic \citep[GalWCat19;][]{Abdullah2020,Abdullah2023} and photometric \citep[redMaPPer;][]{Costanzi2019} SDSS observations, considering all galaxies brighter than $r_{\rm mag} < -20.4$ and within a projected distance $\dsathost < \rvir$ as satellites (right panel; see text for details).
    The TNG-Cluster mass-richness relation is qualitatively consistent with observations.
    }
    \label{fig:halorichness}
\end{figure*}

\section{Methods} \label{sec:meth}

\subsection{TNG-Cluster} 
\label{sec:meth_tng}

TNG-Cluster\footnote{\url{www.tng-project.org/cluster/}} is a suite of 352 massive galaxy cluster simulations, spanning halo masses $\mvir\approx10^{14.3-15.4}\,\msun$ \citep[][\textcolor{blue}{Pillepich et al. in prep.}]{Nelson2024}. These halos were chosen from a $\approx1$~Gpc box-size parent dark matter only simulation, TNG-Cluster-Dark. The 352 halos chosen for re-simulation are based only on $z=0$ halo mass such that: (i) all $\sim 100$ halos with mass $>10^{15}\,\msun$ are included; and (ii) halos with mass $10^{14.3-15.0}\,\msun$ were randomly selected in bins of 0.1~dex such that the halo mass distribution is flat over this mass range \citep[see][for details]{Nelson2024}.

The TNG-Cluster simulation employs the well-tested TNG galaxy formation model \citep{Weinberger2017,Pillepich2018b}. The baryon mass resolution of TNG-Cluster is $m_{\rm bar} = 1.1\times10^{7}\msun$, the same resolution as TNG300 from the original TNG simulation suite \citep{Pillepich2018,Nelson2018,Naiman2018,Marinacci2018,Springel2018}. We note that the TNG galaxy formation model at the TNG-Cluster mass resolution (the same as TNG300) has already been at least partially validated in the low-mass cluster regime \citep[e.g.,][]{Nelson2018,Donnari2021a,Truong2020,Donnari2021b}. Here we briefly summarize the model.  

The TNG simulations, including TNG-Cluster, evolve gas, cold dark matter, stars, and super massive black holes (SMBHs) within an expanding universe, based on a self-gravity + magneto-hydrodynamic framework \citep{Pakmor2011,Pakmor2013} using the \textsc{Arepo} code \citep{Springel2010}. The fluid dynamics employ a dynamic, moving Voronoi tessellation of space. Gas has a cooling floor at $10^{4}$~K, and the relationship between temperature and density for star-forming gas is determined via a two-phase sub-grid pressurization model \citet{Springel2003}. For this analysis, we manually set the temperature of star-forming gas to $10^{3}$~K, its cold-phase value. The TNG galaxy evolution model includes: gas heating and cooling; star formation; stellar population evolution and chemical enrichment from AGB stars and type Ia + II supernovae; supernova driven outflows and winds \citep{Pillepich2018b}; the formation, merging, and growth of SMBHs; and two main SMBH hole feedback modes: a thermal ``quasar'' mode, and a kinetic ``wind'' mode \citep{Weinberger2017}. 

Catalogs contain halos as well as galaxies. Dark matter halos are identified using the Friends-of-Friends (FoF) algorithm with a linking length $b=0.2$, run only using the dark matter particles \citep{Davis1985}. Then the baryonic components are connected to the same halos as their closest dark matter particle. Throughout this paper, we use ``FoF,'' ``group,'' ``FoF group,'' and ``halo'' synonymously. Galaxies are then identified using the \subfind algorithm, which identifies gravitationally bound sets of particles and cells \citep{Springel2001,Dolag2009}. We use the terms ``subhalo'' and ``galaxy'' synonymously even though, in general, \subfind objects may contain no stars and/or gas whatsoever. Typically, the most massive subhalo within a halo is the ``main'' or ``primary subhalo,'' also called the ``central galaxy;'' all other subhalos within a halo are ``satellites.'' We follow the evolution of galaxies using \sublinkgal, which constructs merger trees for subhalos by searching for descendants with common stellar particles and star-forming gas cells \citep{Rodriguez-Gomez2015}. The TNG simulations are publicly available in their entirety, and TNG-Cluster will likewise be released in 2024 \citep{Nelson2019,Nelson2024}. Our analyses adopt the same $\Lambda$CDM cosmology as TNG, consistent with the \citet{Planck2016} results: $\Omega_{\Lambda,0} = 0.6911, \Omega_{\rm m,0} = \Omega_{\rm bar,0} + \Omega_{\rm dm,0} = 0.3089, \Omega_{\rm bar,0} = 0.0486, \sigma_8 = 0.8159, n_s = 0.9667, {\rm and}\ h = H_{\rm 0} / (100\, {\rm km\, s^{-1}\, Mpc^{-1}}) = 0.6774$, where $H_0$ is the Hubble parameter.

\subsection{Halo and satellite galaxy sample selection} \label{sec:meth_sample}

In this work, we exclusively focus on the 352 primary zoom targets from the TNG-Cluster simulation\footnote{We note that there are other halos that happen to be within the individual re-simulation regions \citep{Nelson2024}, but we do not include these objects in our analysis.}. Moreover, we only consider satellite galaxies with stellar mass $\mstarsat>10^{9}\,\msun$, corresponding to $\gtrsim100$ stellar particles each. 

As is common practice with TNG, we define galaxy stellar mass $\mstar$ as the total \subfind stellar mass within twice the stellar half mass radius $\mstar \equiv \mstar(<2\rhalfstar)$. For galaxy gas mass $\mgas$, we take the total gravitationally bound \subfind gas mass, regardless of galactic-centric distance. At times, we consider only cold $\leq 10^{4.5}$~K, hot $>10^{4.5}$~K, ISM ($<2\rhalfstar$), or CGM ($>2\rhalfstar$) gas, where all star-forming gas is cold by definition. The median 
(10th, 90th percentiles) gas cell size in the satellite CGM is $5.2\ (3.0, 8.6)$~kpc. Table~\ref{tab:definitions} summarizes all quantities and definitions.

Unless noted otherwise, we consider galaxies as satellites based on their Friends-of-Friends membership, and do not enforce any explicit restriction on the cluster-centric distance. In some analyses, we rather consider satellites as all galaxies within a 2D projected cluster-centric distance, to mimic observational samples (see \S~\ref{sec:discussion_xray} for more details). In all cases, we only consider subhalos of cosmological origin as defined by the SubhaloFlag in \citet{Nelson2019}. All FoF satellite galaxies considered are uncontaminated at $z=0$, meaning that they contain zero dark matter low resolution elements. However, in general, it is possible that low resolution gas, stars, or dark matter are present, particularly at large distances away from clusters, but this is not expected to influence  our results.

\section{The circumgalactic medium of cluster satellites according to TNG-Cluster} \label{sec:results}

\subsection{Clusters and their satellite demographics} \label{sec:results_demographics}

TNG-Cluster provides an unprecedentedly large set of simulated massive cluster galaxies whose basic demographics are consistent with observations of cluster richness. In particular, in Figure~\ref{fig:halorichness} we plot the 2D histogram of all well-resolved satellites with $\mstar>10^9\, \msun$ across the 352 clusters in the $z=0$ satellite stellar $\mstarsat$ - host halo $\mvirhost$ mass plane (main panel). The color shows the number of galaxies in each bin, which span 0.1~dex in each axis, with 1D histograms for host halo and satellite stellar mass on the top and right subpanels, respectively. 

The 352 galaxy clusters spanning 1~dex in host mass contain a total of $\approx90,000$ satellites, covering 3.5~dex in stellar mass. The maximum satellite stellar mass increases with halo mass, as no satellite galaxy may be more massive than its brightest cluster galaxy (BCG) by definition. For example, the BCG stellar mass within a $10^{14.5}\ (10^{15})\, \msun$ cluster is $\approx 10^{12}\ (10^{12.2})\, \msun$. This is, for example, visible in the 2D histogram (central panel) as a triangular region lacking satellites in the upper left corner. Throughout this work, we refer to satellites $\mstarsat > 10^{10}\, \msun$ as massive, and focus primarily on these systems.  

It should be kept in mind that the TNG-Cluster (+TNG300) halo mass function is approximately flat at masses $\mvirhost\approx10^{14.3-15.1}\, \msun$, whereas there is a sharp decrease in the number of halos at the highest-mass end $\mvirhost\approx10^{15.1-15.4}\msun$, where the sample is volume limited \citep[see fig.~1 from][]{Nelson2024}. As a consequence, in TNG-Cluster, the number of satellites per host mass bin increases with halo mass for lower-mass clusters to then decrease again toward the most massive systems in our sample: this is because more massive clusters on average host more satellites, but more massive clusters are also rarer. On the other hand, the TNG-Cluster satellite stellar mass function (top right subpanel) exhibits the typical shape characteristic also of central galaxies, with a slow decrease for $\mstarsat\sim10^{9-10.5}\, \msun$, and a fast drop-off for massive super-$L_\star$ satellites $\mstarsat\gtrsim10^{10.5}\, \msun$ \citep[e.g.,][]{Baldry2012}.

In Fig.~\ref{fig:halorichness} (bottom left panel), we show a theoretical richness-mass relation, that is,the average number of satellites per host above a given stellar mass threshold, as a function of host mass $\mvirhost$. Solid curves represent medians across the cluster sample, with the 16th and 84th percentiles as shaded regions. As expected, the number of satellites per host above any given stellar mass increases with host mass. For example, at a mass $\mvirhost \approx 10^{14.5}\, (10^{15})\, \msun$, each cluster hosts $\approx 100\ (300)$ satellite galaxies of mass $\mstarsat > 10^{9}\, \msun$ (light orange). Also as expected, at a fixed halo mass, the number of galaxies above a given stellar mass threshold increases as the mass threshold decreases. In TNG-Cluster, nearly all clusters host a few extremely massive satellite galaxies $\mstar>10^{11}\,\msun$ (dark orange). 

Finally, in Fig.~\ref{fig:halorichness} (bottom right panel), we qualitatively compare the richness-mass relation predicted by TNG-Cluster with spectroscopic \citep[GalWCat19;][]{Abdullah2020,Abdullah2023} and photometric \citep[redMaPPer;][]{Costanzi2019} SDSS observations. In particular, we plot the number of satellites per host brighter than $r < -20.4$~mag, corresponding to $\sim \mstarsat \gtrsim 10^{10.4}\, \msun$. Here by satellites we mean all galaxies within the high-resolution zoom region of depth $\sim5\rvirhost \sim 7-12$~Mpc and within a projected distance $<\rvirhost$, regardless of the FoF membership. Even though we do not create synthetic observations to match the SDSS data and hence even though these comparisons are at face value, the result is encouraging: the TNG-Cluster richness-mass relation is qualitatively consistent with SDSS \citep[see also][and their fig. 16]{Nelson2024}.

\begin{figure*}
    \includegraphics[width=\textwidth]{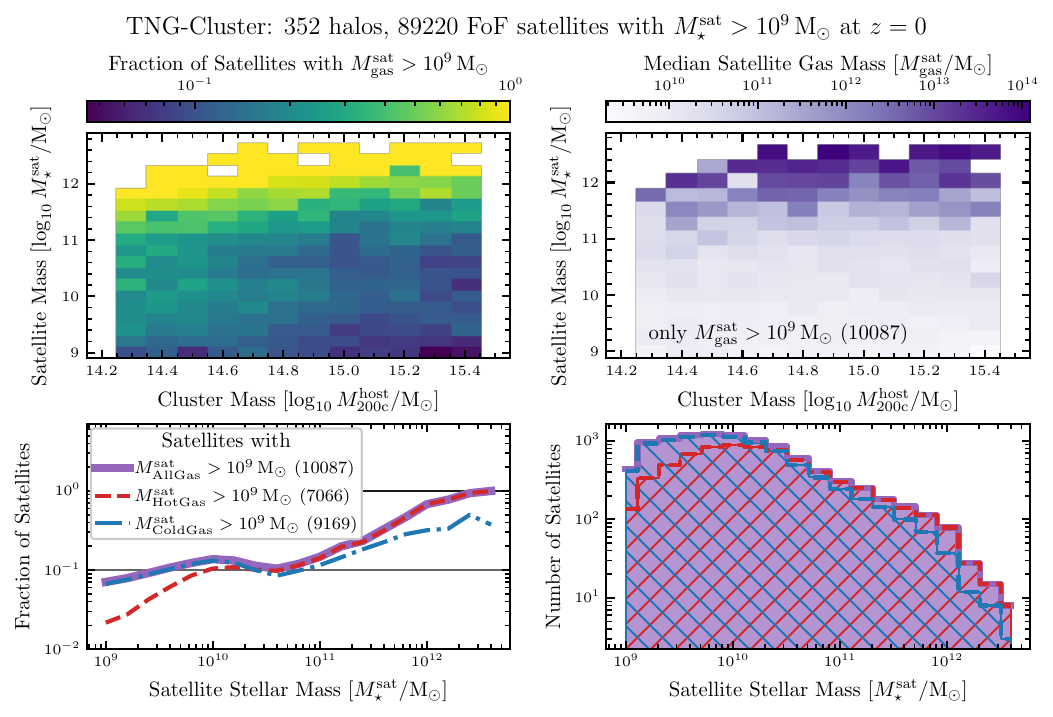}
    \caption{
    {\bf How the TNG-Cluster satellite gas mass varies with the satellite stellar and host halo mass.}
    {\it Top left panel:} for the $\approx90,000$ TNG-Cluster satellites with $\mstarsat>10^{9}\,\msun$, the fraction of satellites that retain gas reservoirs $\mgassat>10^9\,\msun$ today as satellites is shown in bins of satellite stellar $\mstarsat$ and cluster $\mvirhost$ mass. 
    {\it Top right panel:} only $10,000$ (10~percent) of the TNG-Cluster satellites retain gas masses $\mgassat>10^9\, \msun$ today. For these gaseous satellites, we show the median satellite gas mass $\mgassat$ in bins of satellite stellar $\mstarsat$ and host halo $\mvirhost$ mass. 
    {\it Bottom panels:} We show the fractions (left panel) and numbers (right panel) of satellites with all (purple, solid, filled), hot (red, dashed, ``\textbackslash'' hatched), and cold (blue, dashed-dotted, ``/'' hatched) gas masses $>10^9\, \msun$ within a given stellar mass bin at $z=0$. We mark 100~percent and the global average of 10~percent with black lines, and include the total number of gaseous satellites in the legend.
    While a given satellite is more likely to retain gas if it has a higher stellar mass, a given gaseous satellite is more likely to have a lower stellar mass because there are simply more lower-mass satellites.
    }
\label{fig:gasretention}
\end{figure*}

\subsection{The gas content of cluster satellites}
\label{sec:results_gascontent}

Of the $\approx90,000$ TNG-Cluster satellites with $\mstarsat>10^9\msun$, only $\approx10,000$ (10~percent) retain at least some gas today. Throughout the paper, by this we mean at least $\mgassat>10^9\,\msun$ of gravitationally bound gas reservoirs, that is,a well resolved amount of gas. We motivate this choice in Appendix~\ref{app:resolution} and, we have checked that adding an additional criterion on the gas mass fraction $f_{\rm gas} \equiv \mgassat / (\mgassat + \mstarsat) > 10^{-2},\ 10^{-1}$ would remove an unimportant subset of galaxies, that is,40 (0.4~percent), 777 (7.7~percent) of the gaseous satellites.

In Fig.~\ref{fig:gasretention} (top left panel), we explore how this fraction of gas-rich satellites varies with satellite stellar $\mstarsat$ and host cluster $\mvirhost$ mass. At a fixed satellite stellar mass (a given row), the fraction of gas-rich satellites tends to decrease with increasing host mass, broadly consistent with expectations from RPS. Namely, ram pressure increases with host mass, thereby driving down the fraction of gas-rich satellites. Moreover, at a fixed halo mass (a given column), the fraction of gas-rich satellites increases with satellite stellar mass, and this trend exists at all considered stellar masses. The restorative gravitational pull from the stellar body acts as the primary foil to ram pressure, thereby increasing the fraction of gas-rich satellites with increasing satellite mass \citep[e.g.,][]{Wright2022,Rohr2023,Kulier2023}.

In Fig.~\ref{fig:gasretention} (top right panel), we further show that, for the $\approx10,000$ gas-rich satellites, the average satellite gas mass generally increases with satellite stellar mass. In fact, this applies to the most massive satellites and is not the case for those with mass $\mstarsat\sim10^{9-11}\,\msun$, which all exhibit a rather uniform $\mgassat\lesssim10^{10}\,\msun$ of gas reservoir\footnote{For these low-mass satellites, we have checked that the infall look-back time -- that is,the total time the galaxy has been suffering from environmental effects such as ram pressure -- primarily determines how much gas the satellites retain (not shown).}. Indeed, about 100 TNG-Cluster satellites retain gas masses $\mgassat \sim 10^{13-14}\,\msun$ today: these are the most massive satellites in the simulation, and would be in general some of the most massive galaxies in the Universe. We note that these could in fact be considered merging sub clusters \citep[see the companion paper by][]{Lehle2024}. Interestingly, for all gas-rich satellites, the average gas mass does not depend on host mass across the TNG-Cluster mass range. We speculate two origins for the null-trend of median (gas-rich) satellite gas mass with cluster mass. First, and only applying to satellites of mass $\mstarsat\sim10^{9-11}\, \msun$, the average satellite gas mass $\mgassat\sim10^{9-10}\msun$ is just above our threshold for being considered gas-rich $\mgassat > 10^9\, \msun$ (see Appendix~\ref{app:resolution} for more details). Many of these satellites are likely undergoing environmental processes en route toward becoming gas-poor; they have not yet been cluster-members long enough to have been stripped of their gas. Second and related, these satellites may have had their gas reservoirs first preprocessed by other groups before falling into their current cluster hosts \citep[e.g.,][]{Jung2018,Donnari2021a}.

With this intuition that gas retention is primarily determined by satellite mass, in Fig.~\ref{fig:gasretention} (bottom panels), we show how the fraction (bottom left panel) and number (bottom right panel) of satellites that retain significant gas reservoirs at $z=0$ vary with satellite stellar mass. We split the gas (purple, solid, filled) into hot (red, dashed, '/' hatch) and cold (blue, dash-dotted, '\textbackslash' hash) phases. Across all stellar masses considered, $\approx91$~percent of the gaseous satellites retain at least some cold gas. For the hot gas, this only applies to $\approx70$~percent of the gaseous satellites. 

\begin{figure*}
    \begin{subfigure}[b]{\textwidth}
        \includegraphics[width=\textwidth]{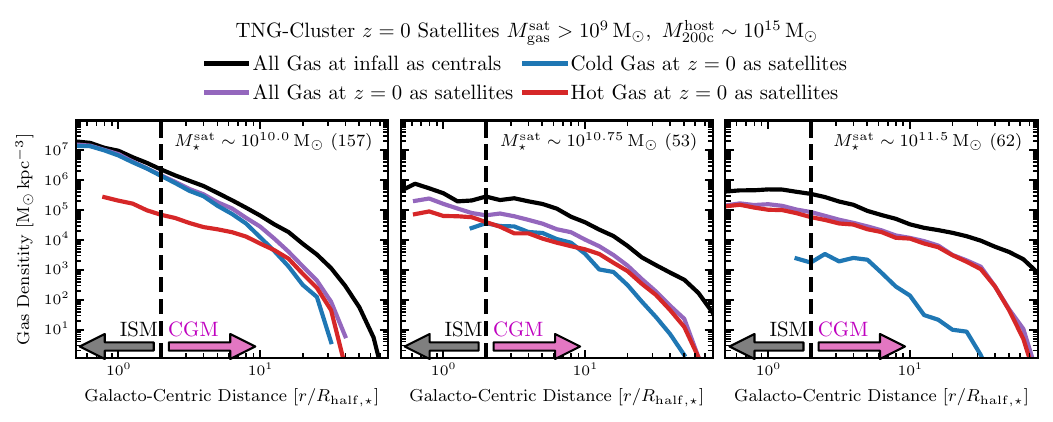}
    \end{subfigure}
    \centering
    \begin{subfigure}[b]{0.666\textwidth}
        \includegraphics[width=\textwidth]{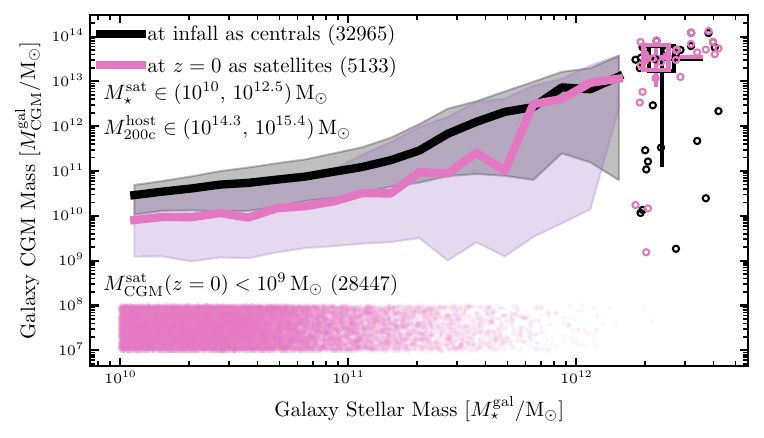}
    \end{subfigure}
    \caption{
    {\bf Spatial extent of satellite gas in TNG-Cluster.}
    Of the $\approx5,000$ gas-rich $\mgassat > 10^9\, \msun$ massive satellites $\mstarsat>10^{10}\, \msun$ at $z=0$, we examine the spatial extent of all (hot and cold) gas at a fixed host cluster mass $\mvirhost\sim10^{15}\,\msun$ and at three stellar masses $\mstarsat\sim10^{10},\, 10^{10.75},\, 10^{11.5}\, \msun$ (top panels, left to right respectively). Moreover, we compare the total gas density (purple) at $z=0$ of cluster members as satellites with that at infall as centrals (black). All curves show the median profiles of all galaxies in the bin; see Fig.~\ref{fig:SGRP_narrowbins_multipanel} for the individual profiles to see the galaxy-to-galaxy variation. We define all gas outside $>2\rhalfstar$ as the  circumgalactic medium (CGM). We then plot total CGM mass $\mcgmsat$ as a function of galaxy stellar mass $\mstarsat$ (bottom panel) at infall as centrals (black), and at $z=0$ as satellites (pink). The curves are the medians of the CGM masses (only the galaxies with nonzero CGM mass for the satellites), and for the most massive galaxies we plot the individual objects (circles) and the median stacks (squares). We manually place the $\approx28,000$ CGM-poor $\mcgmsat < 10^{9}\, \msun$ satellites at masses $\sim10^{7-8}\, \msun$. All stacks use stellar, host, and gas masses at $z=0$, and the radial profiles are normalized to $\rhalfstar$ at the considered time (infall or today). More than 5,000 cluster satellites retain a CGM despite residing in harsh cluster environments.
    }
    \label{fig:satelliteCGM}
\end{figure*}

At low stellar masses $\mstarsat\sim10^{9-10}\,\msun$ the fraction (left) and number (right) of gas-rich satellites increases with stellar mass for all, hot, and cold gas. Nearly all gas-rich satellites retain cold gas at these stellar masses, and the fraction of gas-rich satellites that retain hot gas increases rapidly with stellar mass. This is consistent with ram pressure removing hot, spatially extended gas preferentially first, while the cold ISM gas tends to be more resistant to these environmental effects \citep[e.g.,][Fig.~\ref{fig:satelliteCGM}]{Wright2022,Rohr2023}. 

At intermediate stellar masses $\mstarsat\sim10^{10-10.5}\,\msun$, the fraction of gas-rich satellites (left panel) slightly decreases from $\approx 16$ to $\approx 13$~percent, where the fraction with cold gas (blue, dashed-dotted) drops to $\approx 12$~percent. The decrease in gas retention at these stellar masses is likely due to the onset of the kinetic-mode of AGN feedback \citep[][see below:  Figs.~\ref{fig:satelliteCGM},~\ref{fig:SGRP_narrowbins_multipanel} right panels]{Weinberger2017, Nelson2018, Zinger2020, Truong2020,Ayromlou2021a}. At high stellar masses $\mstarsat\sim10^{10.5-11.75}\,\msun$, the fraction of (cold) gaseous satellites increases with stellar mass to a maximum of $\approx 60$~percent ($\approx 30$~percent). More massive galaxies retain more gas as they (i) have deeper gravitational potential wells; and (ii) tend to be later infallers (into any host; see Table~\ref{tab:definitions} for details), thereby decreasing the amount of time they have been experiencing environmental effects.

While the fraction of gaseous satellites generally increases with mass (top panel), the total number of gaseous satellites decreases with stellar mass (bottom panel), because that the total number of satellites decreases with stellar mass (Fig.~\ref{fig:halorichness}). Thus, while a given satellite is more likely to retain gas if it has a higher stellar mass, a given gas-rich satellite is more likely to have a lower stellar mass, because there are in general many more low-mass satellites.

\subsection{The spatial extent of the satellite gas: the case for the satellite circumgalactic medium} \label{sec:results_satelliteCGM}

To assess the extent of and understand how various physical processes reshape the multiphase gas reservoirs in and around satellites, we turn to the gas radial profiles of the $\approx 10,000$ gas-rich satellites in TNG-Cluster. In Fig.~\ref{fig:satelliteCGM} (top panels), we show the median gas radial profiles (see Fig.~\ref{fig:SGRP_narrowbins_multipanel} for the individual profiles to see the galaxy-to-galaxy variation and Fig.~\ref{fig:SGRP_evolution_example} for the time evolution for an example satellite) separated into all (purple), hot (red), and cold (blue) phases at $z=0$ as satellites at a fixed host cluster mass $\mvirhost\sim10^{15}\, \msun$ and at fixed satellite stellar masses $\mstarsat\sim10^{10},\, 10^{10.75},\, 10^{11.5}\, \msun$ (left to right, respectively). All stacks and $\mstarsat$ measurements are at $z=0$. For comparison, we include the total gas profile at infall when the galaxies were centrals (black). We consider all gas within $2\rhalfstar$ as the interstellar medium (ISM) and all gas at $>2\rhalfstar$ as the circumgalactic medium (CGM).

Across all considered stellar masses and at all galactic-centric distances, galaxies have higher gas densities at infall as centrals, compared to today as satellites. As a result, satellites have lower gas masses and higher quenched fractions than their central counterparts \citep[e.g.,][for TNG]{Stevens2019, Donnari2021a}. In detail, the differences between the gas density at infall and $z=0$ vary both with galactic-centric distance and satellite stellar mass. Moreover, the most prevalent gas phase (hot, red; or cold, blue) today also depends on these two quantities. Complicating the issue, at higher stellar masses $\mstarsat\gtrsim10^{10.5}\, \msun$, nearly all galaxies in the TNG model experience strong AGN kinetic-mode feedback. That is, high-mass satellites are subject to both external and internal processes that impact their gas reservoirs.

At lower stellar masses $\mstarsat\sim10^{10}\, \msun$ (left panel), the total gas density today is composed primarily of cold gas, especially within the ISM. Only at large distances $\gtrsim 5\rhalfstar$ does the hot gas begin to contribute significantly to the total gas density. Within the ISM, the total gas density is largely unchanged for central versus satellite status. However, at the outskirts $\gtrsim10\rhalfstar$, there is a clear truncation in gas profiles. These differences are consistent with expectations from ram pressure removal, where the spatially extended gas is removed preferentially earlier than the tightly bound ISM \citep{Balogh2000}. 

At intermediate stellar masses $\mstarsat\sim10^{10.75}\,\msun$ (middle panel), the total gas density today is composed of similar amounts of hot and cold gas at all distances. Unlike lower-mass galaxies, these have experienced episodes of AGN feedback which partially ejects nearby, mostly cold, ISM gas \citep{Nelson2019b}, reducing its density. At large distances in the CGM $>2\rhalfstar$, the total gas density both at infall and $z=0$ extend to farther distances than at smaller stellar masses. The higher-mass satellites retain sizeable gas reservoirs at large distances up to $\gtrsim 30\rhalfstar$. 

This trend continues to the highest stellar mass bin $\mstarsat\sim10^{11.5}\,\msun$ (right panel). Here, the gas density is comprised almost entirely of hot gas at all distances. In fact, there is little to no cold gas within the ISM, that is,galaxies at these masses are largely quenched, regardless of central or satellite status \citep{Donnari2021a}. The differences between the total gas density profiles as centrals versus as satellites are roughly constant at all distances. The CGM density is highest for these high-mass galaxies, and this satellite CGM gas extends to larger physical distances.

In the bottom panel of Fig.~\ref{fig:satelliteCGM}, we provide one of the key quantitative findings of this analysis: the CGM mass $\mcgm^{\rm gal}$ of satellites as a function of stellar mass $\mstarsat$ (pink), and at infall as centrals (black). The curves are the medians of the CGM masses (only the galaxies with nonzero CGM mass for the satellites), and for the most massive galaxies we plot the individual objects (circles) and the median stacks (squares). Here by CGM we mean the gas reservoir that extends beyond the main stellar body of a galaxy. According to TNG-Cluster, more than 5,000 massive $\mstarsat>10^{10}\, \msun$ satellites retain their own CGM or gaseous atmosphere despite the harsh cluster environments. Further, on average, the retained CGM mass increases with satellite stellar mass, while the differences between infall and $z=0$ decrease. higher-mass satellites are more resistant to environmental effects and are able to retain spatially extended gas reservoirs. Beyond the average trends, however, there remains a large galaxy-to-galaxy variation in the ability of retaining CGM mass, and we study exactly this diversity next.

\subsection{The diversity of satellite circumgalactic media} \label{sec:results_diversity}

\begin{figure*}
    \includegraphics[width=\textwidth]{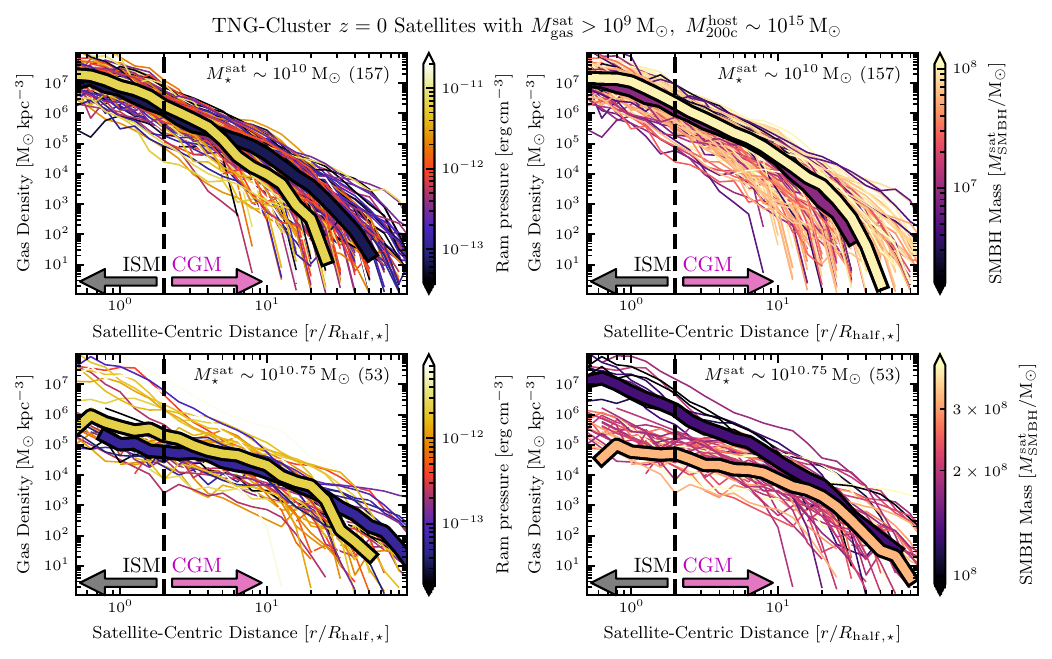}
    \caption{
    {\bf How ram pressure and SMBHs affect the gas contents of TNG-Cluster satellite galaxies.}
    For the gas-rich $\mgassat>10^{9}\, \msun$ satellites at a fixed cluster $\mvirhost\sim10^{15}\,\msun$ and satellite $\mstarsat\sim10^{10},\, 10^{10.75}\, \msun$ (top versus bottom panels) mass, we show total gas density radial profiles colored by current ram pressure (left column) and SMBH mass (right column). The thick curves show the medians of the top and bottom quartiles of ram pressure and SMBH mass, and the thin curves are the profiles of the individual satellites. 
    Ram pressure tends to truncate the gas profiles at large distances and slightly compress the ISM gas, while SMBH feedback tends to eject and decrease the density of the ISM gas for more massive satellites. 
    }
\label{fig:SGRP_narrowbins_multipanel}
\end{figure*}

\begin{figure}
    \includegraphics[width=\columnwidth]{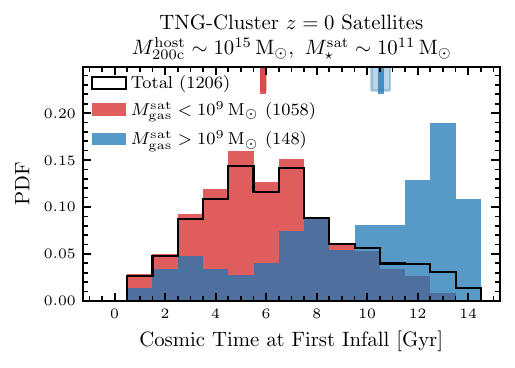}   
    \caption{
    {\bf Gas-rich satellites today are typically late infallers.}
    We plot the probability distribution functions (PDFs) of the infall times of all (black), gas-poor $\mgassat<10^{9}\, \msun$ (red), and gas-rich $\mgassat>10^{9}\, \msun$ (blue) TNG-Cluster satellites at a fixed host halo $\mvirhost\sim10^{15}\, \msun$ and satellite stellar $\mstarsat\sim10^{11}\, \msun$ mass. We mark the medians and errors of the distributions as hashes on the top $x$-axis. The infall time is the first infall into any host, regardless if the galaxy has been preprocessed or not. At a fixed host halo and satellite stellar mass, the infall time is the primary driver determining whether satellites remain gas-rich or gas-poor today: more recent infallers are more likely to retain large gas reservoirs. 
    }
    \label{fig:infall-time_countors}
\end{figure}

\begin{figure}
    \includegraphics[width=\columnwidth]{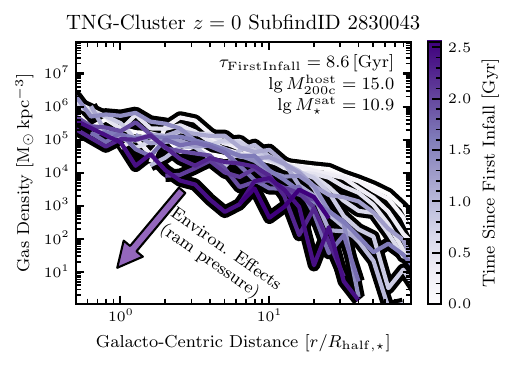}
    \caption{
    {\bf How the environment removes gas from an example satellite.}
    For an example gas-poor satellite today of mass $\mstarsat(z=0) \approx 10^{10.9}\, \msun$ in a cluster of mass $\mvirhost(z=0) \approx 10^{15.0}\, \msun$, we study the time evolution of the gas radial profiles since first infall ($\tau_{\rm FirstInfall} \approx 8.6$~Gyr i.e., look-back time $\approx 5$~Gyr) until it becomes gas-poor today. The cumulative environmental effects, namely the integrated ram pressure since infall, and secular processes, namely feedback from the SMBH, remove the satellite's gas, leaving a gas-poor, quenched satellite today.
    }
    \label{fig:SGRP_evolution_example}
\end{figure}

We now consider individual galaxies to explore the diversity of satellite CGM. In Fig.~\ref{fig:SGRP_narrowbins_multipanel} we plot the total gas density radial profiles for all gas-rich $\mgassat>10^{9}\, \msun$ satellites at a fixed host halo $\mvirhost\sim10^{15}\, \msun$ and satellite stellar mass $\mstarsat\sim10^{10},\ 10^{10.75}$. We color the profiles by their instantaneous ram pressure\footnote{We compute the instantaneous ram pressure by measuring the local background environment using an adaptive spherical shell and Gaussian mixture estimator \citep{Ayromlou2019,Ayromlou2021b}. The ram pressure is then the background density multiplied by the relative velocity squared.} (left column) and supermassive black hole (SMBH) mass (right column), and we overplot the medians of the top and bottom quartiles of ram pressure and SMBH mass. We also examine the effect of time since infall, a proxy for the integrated environmental effects, on the radial profiles and find similar results that of the instantaneous ram pressure (not shown).

At lower stellar masses $\mstarsat\sim10^{10}\,\msun$, satellites experiencing more ram pressure have lower CGM densities, and their spatial extent is truncated (top left panel). This result also holds for higher satellite masses (lower left panel). In the ISM $<2\rhalfstar$, gas densities tend to increase with ram pressure, but the effect is small. At higher stellar masses $\mstarsat\sim10^{10.75}\,\msun$ (bottom left panel), this effect is stronger. Recent simulations and radio observations predict and infer that ram pressure can compress a satellite's ISM \citep[e.g.,][]{Vulcani2018,Roberts2022,Kulier2023}, and our results qualitatively agree.

To examine the effect of SMBH feedback we use SMBH mass as a proxy. This is the integral of the SMBH accretion history, and so it is a proxy of and proportional to the total feedback energy ever released. At lower stellar masses $\mstarsat\sim10^{10}\,\msun$ (top left panel), SMBHs do not seem to have an impact on satellite gas densities. Within the TNG model, the SMBHs in these galaxies are largely in the quasar mode, where the corresponding feedback couples poorly to the gas, and is thus not ejected \citep{Weinberger2018}. 

At higher stellar masses $\mstarsat\sim10^{10.75}\,\msun$ (bottom right panel), SMBHs are more important. At these masses and by $z=0$, most of these satellites have experienced episodes of kinetic AGN feedback. Those with lower SMBH masses have higher ISM densities by $\sim1-2$~dex than those with higher-masses. This effect extends into the CGM, to $\sim10\rhalfstar$. When SMBHs begin accreting at low rates, exerting wind mode feedback and expelling the ISM gas, they partially remove their own source of fuel for gas accretion. Without a supply of nearby cold gas, SMBHs grow more slowly. As a result, finding SMBHs in satellites that are much more massive is rare, although in central galaxies they can continue to grow via mergers \citep{Joshi2020}. At large radii $\gtrsim10\rhalfstar$ there is little impact from SMBH feedback. So, in summary, SMBH feedback tends to eject and decrease the density of the ISM gas for more massive, gas-rich satellites $\mstar\gtrsim10^{11}\,\msun$, which are typically quenched as a result but can still hold on to their CGM.

Now as seen in the previous sections, while many massive satellites manage to retain their own CGM today, at stellar masses $\mstarsat\lesssim10^{11.5}\, \msun$ a majority of TNG-Cluster satellites have actually lost their gas reservoirs due to combinations of secular and environmental processes. Moreover, among the gas-rich massive satellites, the retained CGM mass can vary by 2 or 3 orders of magnitude (Fig.~\ref{fig:satelliteCGM}, main panel).

For high-mass satellites $\mstarsat\sim10^{11}\,\msun$, we therefore study why most ($\sim80$~percent; Fig.~\ref{fig:gasretention}, bottom left panel) end up gas-poor at $z=0$. Fig.~\ref{fig:infall-time_countors} shows the probability distribution functions (PDFs) of infall times for satellites at a fixed host halo $\mvirhost\sim10^{15}\, \msun$ and satellite stellar $\mstarsat\sim10^{11}\, \msun$ (total, black), split into those that are gas-poor $\mgassat < 10^{9}\, \msun$ (red), and those that are gas-rich $\mgassat>10^{9}\, \msun$ (blue) today. The infall time is the first infall into any host, regardless if the galaxy has been preprocessed or not. Gas-rich satellites tend to be late-infallers, experiencing environmental effects for a shorter period of time than gas-poor satellites. 

We further explore the cumulative environmental effects in Fig.~\ref{fig:SGRP_evolution_example}, where we show the time-evolution since infall ($\tau_{\rm FirstInfall} \approx 8.6$~Gyr, i.e., look-back time $\approx 5$~Gyr ago) of the gas radial profiles for an example gas-poor satellite today ($\mstarsat(z=0)\sim^{11}\msun$ in a cluster of mass $\mvirhost(z=0)\sim10^{15}\,\msun$). As the time since infall increases, the gas density at a fixed galacto-centric distance decreases, and the maximum extent of the CGM truncates, approximately following the direction of arrow. After sufficient time in the cluster environment, subject to ram pressure from its passage through the ICM, the galaxy loses all of its gas, becoming a gas-poor, quenched satellite today.

\section{Implications of satellite CGM in clusters} \label{sec:discussion}

As we have seen in the previous sections, massive satellite galaxies can retain a spatially extended, mostly hot CGM, and each cluster hosts a few to tens such massive satellites. One would then expect the satellite CGM to emit in the X-ray and to contribute to the total X-ray flux from galaxy clusters. Moreover, satellite CGM could contribute to absorption lines on background quasar spectra in the near UV. We thereby study the possibility to statistically detect this satellite CGM in observations.

These results and the following discussion depend on a number of assumptions: the TNG galaxy formation model, the Friends-of-Friends halo finder for defining the halos and their members, the \subfind algorithm for determining the galaxies' bound resolution elements, the TNG-Cluster halo sample, and the numerical resolution of TNG-Cluster itself (see Appendix~\ref{app:resolution} for more details on the gas resolution).

\subsection{Detecting extended soft X-ray emission around satellites} \label{sec:discussion_xray}

\begin{figure*}
    \includegraphics[width=0.99\textwidth]{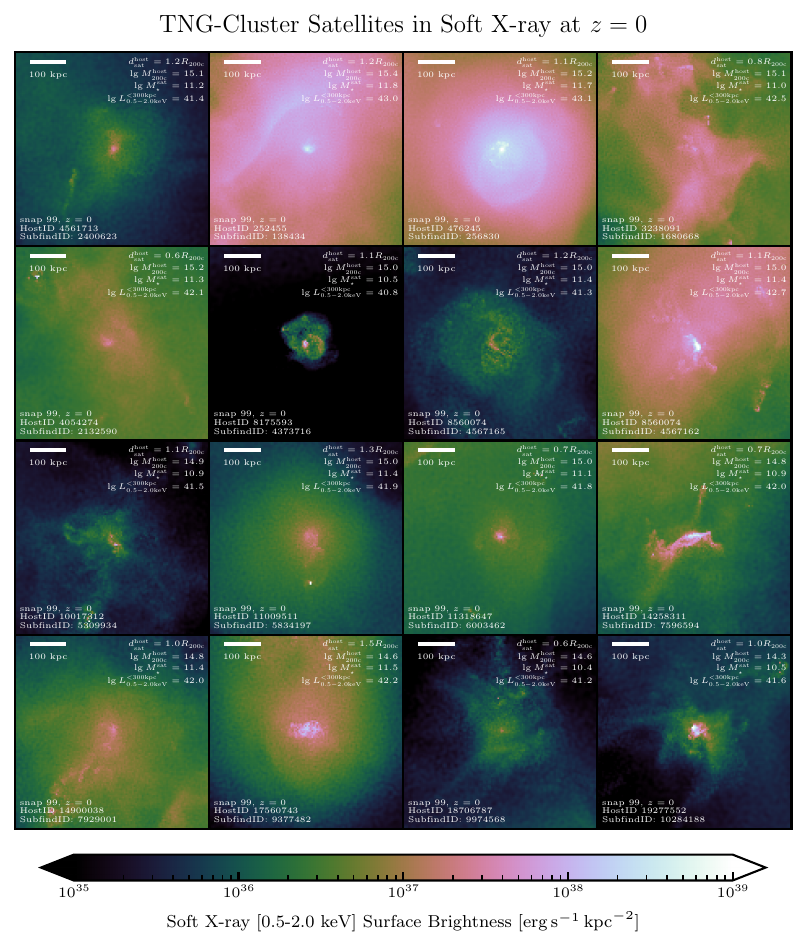}
    \caption{
    {\bf Poster of the diversity of TNG-Cluster satellites in soft X-ray emission.}
    We show the soft X-ray (0.5-2.0~keV) surface brightness for 16 example TNG-Cluster satellite galaxies at $z=0$. Each image is 600$\times$600~kpc$^2$ in size (scale bar in the upper left). These galaxies are necessarily within a projected distance of $1.5\rvir$ of their host. We include information about the galaxy and its host in the upper right and lower left corners; the units of the cluster $\mvirhost$ and satellite $\mstarsat$ mass are $[\log_{10}\, \msun]$, and the soft X-ray luminosity $L_{\rm 0.5-2.0keV}^{\rm <300kpc}$ is the total X-ray luminosity within a projected aperture of 300~kpc in units of $[\log_{10}\, {\rm erg\, s^{-1}}]$. Many of the maps display morphological signatures of SMBH feedback, with dome-like inflated bubbles and/or ram pressure, with gas tails extending in specific directions.
    }
    \label{fig:xray_poster}
\end{figure*}

We now turn to answer the following question: Can the hot CGM retained by satellites be detected, above the background ICM \citep{Schuecker2001}. This hot satellite gas emits thermally via bremsstrahlung in the X-ray, similarly to the ICM. However the gas temperature is much lower for satellite CGM gas $\approx0.5-2$~keV compared to the ICM at $\approx5-10$~keV \citep[see also][]{Truong2024}. We therefore focus on the soft X-ray $0.5-2$~keV emission, computed using the collisional ionization tables of AtomDB from the Astrophysical Plasma Emission Code \citep[][following \citealt{Nelson2023}]{Smith2001}. 

In Fig.~\ref{fig:xray_poster} we show examples of broad-band soft X-ray emission around 16 TNG-Cluster satellite galaxies. These galaxies are necessarily within a projected distanced $<1.5\rvirhost$ of one of the 352 TNG-Cluster BCGs. Each image is $600\times600$~kpc$^2$ in size, and we include information about each galaxy and its host in the panels. The X-ray emission around these galaxies is clearly visible against the background ICM, even without a background subtraction. The X-ray morphologies are diverse; while many examples are roughly circular, many also show asymmetries and filamentary features. These morphologies could be caused by both internal (for example, SMBH feedback, such as for SubfindIDs 256830, 4373716, 7596594) and/or external processes (for example, ram pressure, such as for SubfindIDs 7596594, 7929001). These are some of the most prominent spatially extended X-ray emission around satellites, but each cluster hosts several such examples\footnote{We include three projections of each cluster in the soft X-ray in an online gallery, where satellites can clearly be seen against the background: \url{https:/www.tng-project.org/cluster/gallery/}.} \citep{Nelson2024}. 

\begin{figure*}
    \includegraphics[width=\textwidth]{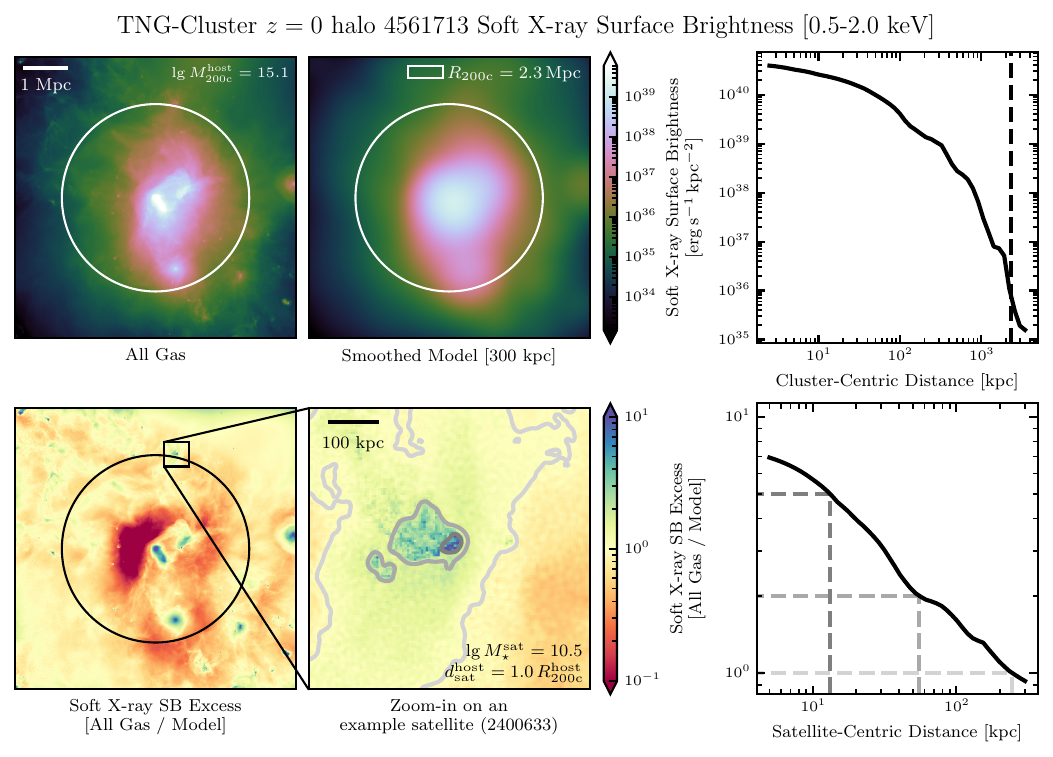}
    \caption{
    {\bf Schematic detailing how we detected satellites' circumgalactic media and measured their radial profiles.}
    For a given halo (halo mass $\mvirhost\approx10^{15.1}\, \msun$) at $z=0$, we compute the soft X-ray ($0.5-2.0$~keV) continuum surface brightness (SB) on $5\times5\, {\rm kpc^{2}}$ pixels (top left). We then smooth the image using a Gaussian kernel of standard deviation 300~kpc (top center). The white circles indicate the virial radius. We include the azimuthally averaged radial profile, where the dashed line marks the virial radius (top right). We compute the soft X-ray surface brightness excess as the ratio of all the gas in the image to the smoothed model (bottom left), and zoom in on an individual example satellite (bottom center; stellar mass $\mstarsat\approx10^{10.5}\, \msun$ at projected cluster-centric distance $\dsathost\approx1.0\rvirhost$). We include the contours of \{1, 2, 5\} times the background on the zoom-in and on the satellite's excess radial profile (bottom right). In this satellite we detect a soft X-ray surface brightness excess out to $\approx200$~kpc from the satellite's center, and the morphology suggests the satellite's X-ray emitting gas is experiencing environmental effects. The soft X-ray excess profile for this example satellite is similar to others of the same stellar mass and cluster-centric distance (see Fig.~\ref{fig:Lxsoft_stacks} and text).
    } 
    \label{fig:soft-xray-excess-multipanel}
\end{figure*}

We now quantify the spatial extent of the X-ray emitting gas compared to the ambient background for all $>30,000$ massive satellite galaxies around TNG-Cluster hosts. Fig.~\ref{fig:soft-xray-excess-multipanel} displays a schematic of our methodology. We start with the projected soft X-ray map of an entire cluster (top left panel; this example has mass $\mvirhost\approx10^{15.1}\, \msun$), where the image is $3\rvir$ ($\approx6.9$~Mpc) in size in each direction and the white circle marks $\rvir$. We include all gas in the simulation in this projection, with a depth of $\sim5\rvir\sim10$~Mpc. We then smooth the X-ray map using a Gaussian kernel with a fixed physical width of 300~kpc (top center panel). This choice of smoothing length enables us to detect excesses and deficits compared to the background on scales smaller than 300~kpc. On larger scales, the signal is smoothed out and tends toward the background medium. We vary this smoothing scale between 100 and 1,000~kpc, finding qualitatively similar results. The projected soft X-ray surface brightness radial profile (top right panel) decreases with halo-centric radius, where the dashed line marks the halo radius. This radial profile itself does not provide a good background estimate, as the ICM can be far from spherically symmetric \citep[e.g.,][]{Truong2021}.

\begin{figure*}
    \includegraphics[width=\textwidth]{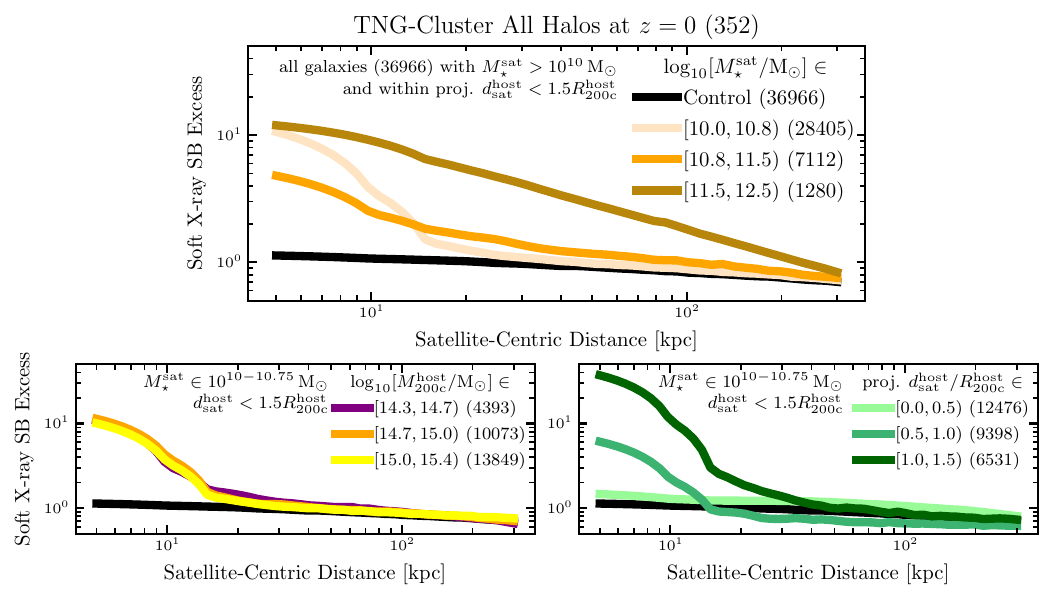}
    \caption{
    {\bf Stack of the $z=0$ satellite soft X-ray surface brightness excess radial profiles around all 352 TNG-Cluster halos.}
    We compute the soft X-ray (0.5-2.0~keV) surface brightness (SB) excess around all $\approx37,000$ galaxies with $\mstarsat>10^{10}\, \msun$ and within a projected distance $\dsathost<1.5\rvirhost$ of the 352 hosts. We then mean-stack these radial profiles in bins of satellite stellar mass $\mstarsat$ (top panel), and within a fixed stellar mass bin, we further stack in bins of host cluster mass $\mvirhost$ (bottom left) and cluster-centric distance (bottom right). In all panels, the control sample represents the same measurement, at random locations such that the distributions of cluster-centric distances match that of the satellites. Across all satellites, the X-ray excess extends to $\sim20-100$~kpc, although individual satellites or specific stacks may exhibit excesses out to $\approx300$~kpc from the satellites.
    }
    \label{fig:Lxsoft_stacks}
\end{figure*}

We compute the soft X-ray surface brightness excess map (bottom left) as the normal X-ray map divided by the smoothed model. In this map the perturbations -- both excess (green-blue) and deficits (orange-red) -- are clearly visible. Satellites and their CGM produce many of these perturbations, both point-like and extended excess sources. Here, we zoom in on one example satellite of mass $\mstarsat \approx 10^{10.5}\, \msun$ and at projected host-centric distance $\dsathost\approx 1.0\rvirhost$ (bottom center panel), where the zoom region is 600~kpc per side. The contours mark where the X-ray surface brightness reaches \{1, 2, 5\} times the smoothed background. The region of excess X-ray emission (inside the outermost contour) extends to large distances and has a complex morphology. The inner, brighter regions show signs of ram pressure removal. We show the projected radial profile (lower right panel), and find that the excess X-ray emission decreases with satellite-centric distance, reaching the background value at $\approx200$~kpc. This example X-ray excess profile is similar to the average of all satellites of this mass and distance, although the maximum excess at small satellite-centric distances for this example is lower than average (see text below and Fig.~\ref{fig:Lxsoft_stacks}, bottom right panel).

We next consider all massive galaxies $\mstarsat > 10^{10}\, \msun$ within a projected cluster-centric distance $\dsathost < 1.5\rvirhost$, ignoring the FoF membership. This yields $\approx37,000$ satellites around the 352 clusters. We zoom-in on the positions of each of these satellites and compute the soft X-ray surface brightness excess radial profile. Fig.~\ref{fig:Lxsoft_stacks} shows the mean-stacked profiles in bins of satellite stellar mass $\mstarsat$ (top center), host halo mass $\mvirhost$ (bottom left), and cluster-centric distance (bottom right). As a reminder, $\approx85$~percent of these satellites are gas-poor and do not contribute to the total emission, but we still include them in stacks\footnote{When instead using median-stacks rather than mean-stacks, X-ray excesses around the most massive satellites ($\mstarsat \in 10^{11.5-12.5}\, \msun$) and satellites at the largest cluster-centric distances ($\dsathost / \rvirhost \in 1.0-1.5$) remain, but all other stacks are indistinguishable from the control.} . Some of these galaxies lie in projection near other X-ray excesses (deficits) and would still contribute (negatively) to the stacked profile. It is not clear a priori if the excesses and deficits cancel each other out. We therefore construct a control sample stack that matches the distribution of cluster-centric distances. While we construct control samples for each stack (in satellite mass, host mass, and cluster-centric distance), the differences between the controls of each stack are negligible, and all show a surface brightness excess of $\approx 1$. We simplify by plotting the total control sample (black curves) in each panel.

In Fig.~\ref{fig:Lxsoft_stacks} (top center), all three stellar mass $\mstarsat$ bins display X-ray excesses compared to the control. The most massive satellites $\mstarsat\sim10^{11.5-12.5}\,\msun$ (dark orange), have an excess of nearly $\approx10\times$ the background at small satellite-centric distances $\lesssim10$~kpc, and the excess extends out to $\approx300$~kpc. The intermediate $\mstarsat\sim10^{10.75-11.5}\,\msun$ (orange) and lower $\mstarsat\sim10^{10-10.75}\,\msun$ (light orange) bins also display extended X-ray excesses out to $\approx 100$ and $\approx30$~kpc respectively. The satellites in the lower-mass bin have a higher peak X-ray excess at small distances than those in the intermediate mass bin. As shown in  Fig.~\ref{fig:satelliteCGM} (top panels), the hot gas density at small distances $\lesssim \rhalfstar$ is higher for lower-mass satellites $\mstarsat\sim10^{10}\,\msun$ than for intermediate-mass satellites $\mstarsat\sim10^{10.75}\,\msun$. Using hot gas density as a proxy for soft X-ray luminosity, we would then expect these lower-mass satellites (light orange) to be brighter at small distances than intermediate-mass satellites (orange). We speculate that the lower hot gas density in the centers of high-mass satellites could be caused by the AGN feedback redistributing gas to larger distances \citep[e.g.,][]{Ayromlou2023}.

To distinguish from the effects of satellite mass, we now further stack the X-ray excess radial profiles by host mass $\mvirhost$ (bottom left) and cluster-centric distance $\dsathost$ (bottom right) only for the satellites in the lowest-mass bin $\mstarsat\in10^{10-10.75}\, \msun$. In the X-ray excess radial profiles stacked by host mass $\mvirhost$ (bottom left), all three halo mass bins display similar maximum excess of $\approx10\times$ the background at small distances and similar extents to $\approx50-100$~kpc. There are no significant differences between different host masses, in agreement with our earlier result that the gas mass of gas-rich satellites $\mgassat>10^9\,\msun$ does not vary with halo mass (Fig.~\ref{fig:gasretention}, top right panel).

Lastly, in the stack by projected cluster centric distance $\dsathost$ (bottom right), satellites at larger projected cluster-centric distances have both higher peak X-ray excesses and extend to farther satellite-centric distances. Satellites at projected distances $\dsathost \approx 1.0-1.5\rvirhost$ (dark green) have maximum excesses of $\gtrsim 30\times$ the background and extend out to $\approx 50$~kpc from the satellite. Greater excesses at large distances arise because these satellites retain more of their CGM. They likely consist of mostly first infallers into the clusters, and so have not been experiencing the cluster-environment effects for as long as other satellites. Since the satellites at the closest projected distances $\dsathost \lesssim 0.5\rvirhost$ (light green) are, and have been, experiencing stronger ram pressure removal of their gas, it is natural that they have no significant excess X-ray emission compared to the control. 

According to our theoretical experiment and on the outcome of TNG-Cluster, stacking the soft X-ray excess emission around the positions of optically selected satellites can yield a clear signal of the satellite CGM emission. We expect these results to qualitatively hold in X-ray surveys, and such an experiment could be conducted with the eROSITA all sky survey, albeit only for a sample of nearby clusters \citep[e.g.,][]{Comparat2022,Zhang2024}. Recently, \citet{Hou2024} stack the archival {\it Chandra} observations (0.5-2.0~keV) of 21 star-forming, edge-on, late-type galaxies in the Virgo cluster. They find three detections without the need for stacking, and, when stacking by satellite SFR, they detect a signal for the highest SFR bin, with an X-ray luminosity $L_x \sim 10^{38}\, {\rm erg\, s^{-1}}$ per galaxy. While all of the brightest example TNG-Cluster satellites exceed this luminosity (Fig.~\ref{fig:xray_poster}), we speculate that many of the fainter TNG-Cluster satellites would emit at approximately this luminosity. Finally, although we have focused on the broad-band soft X-ray as an example for satellite CGM detectability, we also expect the CGM to be detectable in SZ, by using, for example, X-ray hardness \citep[e.g.,][]{Truong2021}, and in ratios of X-ray emission lines using XRISM or LEM \citep{Kraft2022}.

\subsection{Implications of satellite CGM for ICM emission studies} \label{sec:discussion_absorption}

\begin{figure*}
    \centering
    \includegraphics[width=0.95\textwidth]{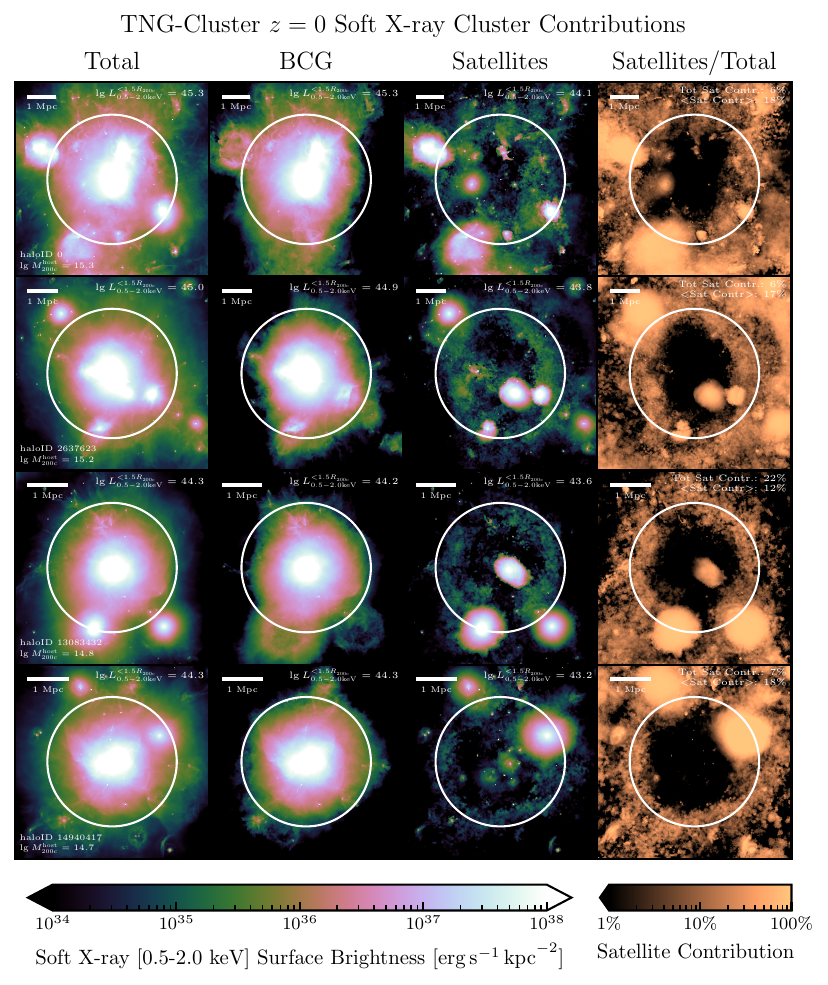}
    \caption{
    {\bf Separating cluster soft X-ray emission into its main sources, for four clusters in TNG-Cluster.}
    We break down the total soft X-ray emission (Total) of an example cluster into its main components: gas that is gravitationally bound to the main halo, i.e., brightest central galaxy (BCG), and gas bound to the satellite galaxies (Satellites). Here, satellites are all galaxies besides the BCG that lie within the field of view of the image $3\rvir\times3\rvir$. We include the haloID and mass $\mvirhost$ in units of $[\log_{10}\, \msun]$ in the lower left in the Total image, and the soft X-ray luminosity $L_{\rm 0.5-2.0keV}^{\rm <1.5\rvir}$ within a projected aperture $1.5\rvir$ in units of $[\log_{10}\, {\rm erg\, s^{-1}}]$ in the top right of each panel. We compute the fractional contribution from satellites to the total soft X-ray emission (Satellites / Total), and include both the total soft X-ray contribution from satellites and the mean satellite contribution per pixel in the upper right corner. In all panels the circles denote $\rvir$. The satellite contributions vary from halo-to-halo, but in general satellites contribute $\sim10$~percent of the total soft X-ray flux within $\lesssim1.5\rvir$ of the clusters. 
    }
    \label{fig:soft-xray_onlySatellites_mosaic}
\end{figure*}

\begin{figure*}
    \includegraphics[width=\textwidth]{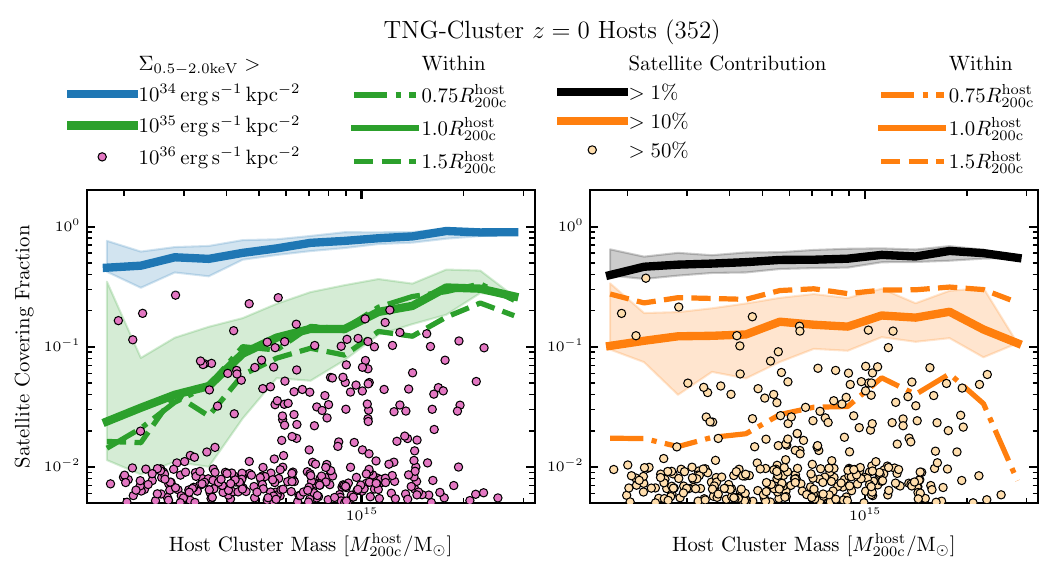}
    \caption{
    {\bf Covering fraction of X-ray emission from satellite circumgalactic media in TNG-Cluster.}
    We measure the projected covering fraction of satellites soft X-ray (0.5-2.0~keV) emission to total for the 352 clusters at $z=0$ for two cases: above a given fixed satellite surface brightness threshold (left panel) and above a given satellite-to-total fractional local threshold (right panel). The fiducial threshold choices are shown for varying cluster-centric apertures. Thick curves are the medians across all clusters, shaded regions enclose the 16th and 84th percentiles, and points are the individual clusters, where we manually place clusters with satellite covering fractions $<10^{-2}$ between $10^{-2}$ and $10^{-2.5}$. In both panels when varying the minimum threshold we use the fiducial aperture of $<1.0\rvir$, and when varying the aperture we use the intermediate threshold ($>10^{35}\, {\rm erg\, s^{-1}\, kpc^{-2}}$ in the left panel; $>10$~percent contribution in the right panel). The collection of satellite CGM can contribute to a significant portion of the cluster's projected area, and at larger surface brightness or satellite-to-total fractional thresholds there is more halo to halo variation. 
    }
    \label{fig:coveringfraction}
\end{figure*}

Considering that each cluster hosts tens to hundreds of massive satellites and that many can retain their own CGM, we turn to the covering fraction of satellite CGM on a cluster by cluster basis. While the overall luminosity and mass fractions of satellite CGM compared to the total ICM may be small, it may be an important component of observable soft X-ray emission from the ICM.

In Fig.~\ref{fig:soft-xray_onlySatellites_mosaic}, we split the total (left column) soft X-ray emission of four example clusters into its two main components: gas gravitationally bound to the main halo or BCG (second column); gas gravitationally bound to satellite galaxies (third column). We consider all galaxies within the projected field of view as satellites, regardless of stellar mass or FoF membership. A majority of the TNG-Cluster soft X-ray emission within $\rvir$ (circles) comes from the gas bound to the BCG, but there are also many satellites with their hot CGM that contribute. We compute and show the fractional contribution from satellites to the total soft X-ray emission (rightmost column). 

In the cluster centers, satellites do not contribute significantly to the total X-ray flux, except for, locally, around the positions of particularly bright satellites. In the outskirts, however, satellites contribute more significantly to the total X-ray. In each cluster here there are at least a few satellites with bright, spatially extended CGM, and all clusters host many satellites with visible CGM emission.

We quantify the projected covering fraction of soft X-ray emission from bound satellite gas in Fig.~\ref{fig:coveringfraction}. We compute the projected covering fraction of satellite soft X-ray (0.5-2.0~keV) emission within a given aperture for the 352 clusters at $z=0$ in two ways: a) above a given surface brightness threshold (left panel) or satellite-to-total fractional threshold -- that is,above a given threshold of local satellite-to-total fractional emission on a pixel by pixel basis (right panel). Thick curves are the medians, shaded regions enclose the 16th and 84th percentiles, and points are the individual clusters.

Above any of the considered physical surface brightness thresholds (left panel, different colors), the satellite covering fractions increase with cluster mass. As the cluster mass increases, the number of satellites and average satellite mass increase, while higher-mass satellites are also more likely to retain larger, brighter CGM. At a surface brightness above $10^{34}\, {\rm erg\, s^{-1}\, kpc^{-2}}$ (blue), nearly all clusters have a satellite covering fraction of $\approx0.7-1$, suggesting that with deep enough X-ray observations a majority of cluster-sightlines intercept some satellite circumgalactic gas. That is, while the current eROSITA all sky survey (eRASS:4) can detect X-ray surface brightness down to $\sim10^{35}\, {\rm erg\, s^{-1}\, kpc^{-2}}$ \citep[at least via stacking;][]{Zhang2024}, we speculate that at $\sim1$~dex deeper imaging more than of pixels would contain emission from satellites. Detecting this satellite contribution would depend on the signal to noise, or using X-ray spectra to distinguish the satellite CGM from the ICM emission.

Above $10^{35}\, {\rm erg\, s^{-1}\, kpc^{-2}}$ (green; the level currently achievable in eRASS:4), the satellite covering fraction increases from $\approx3$~percent at $\mvirhost\sim10^{14.3}\, \msun$ to $\approx30$~percent at $\mvirhost\sim10^{15.4}\, \msun$. Above $10^{36}\, {\rm erg\, s^{-1}\, kpc^{-2}}$ (pink) the median satellite covering fraction is $\lesssim10^{-1}$ for halo masses $\lesssim10^{15}\, \msun$ (not shown). There are however a number of individual clusters at all masses with high covering fractions, that is,$>0.1$ within $\rvirhost$. Finally, at a surface brightness threshold of $>10^{35}\, {\rm erg\, s^{-1}\, kpc^{-2}}$, the covering fractions are similar within $0.75\rvirhost \approx R_{\rm 500c}^{\rm host}$ (dashed-dotted) and $\rvirhost$ (solid), decreasing at larger radii $\approx1.5\rvirhost$ (dashed).

Roughly half the cluster project area within the virial radius has $>1$~percent contribution from satellites (Fig.~\ref{fig:coveringfraction}, black line). Only $\approx10$~percent of the pixels have satellite contributions above the fiducial 10~percent threshold. At the highest satellite-to-total threshold of $>50$~percent, meaning that at least half of the total soft X-ray flux per pixel originates from satellites, only $\approx3$~percent of clusters have significant satellite covering fractions $>10$~percent. As can been seen from the images in Fig.~\ref{fig:soft-xray_onlySatellites_mosaic}, the X-ray emission in cluster cores is dominated by the gas bound to the BCG. In the outskirts however, satellites contribute more significantly to the total flux. Considering the region from $0.75-1.5\, \rvirhost$ ($\approx 1.0-2.0\, R_{\rm 500c}^{\rm host}$), the covering fractions above the fiducial $>10$~percent satellite contribute are the highest at $\approx 0.75$. Overall, satellites contribute $\gtrsim10$~percent of the total soft X-ray emission to cluster outskirts $\approx0.75-1.5\rvirhost$.

\subsection{Implications for satellite galaxy evolution} \label{sec:discussion_evolution}

The presence of a CGM, or not, affects the evolution of a given satellite galaxy and how the satellite interacts with its surrounding medium. From the perspective of the satellites themselves, the presence of an extended CGM gaseous halo and reservoir will modify their evolution in high-density environments, versus if they were directly exposed to the cluster environment \citep{Li2023}. Conversely, from the perspective of the cluster-satellite interactions, feedback from the satellites will have different effects on the surrounding host medium if they have their own CGM \citep{Bahe2012}. Finally, from the perspective of the CGM and/or ICM of massive haloes, whether satellites are surrounded by their own gas will affect the interpretation of the ICM itself also via, for example, absorption line studies \citep{Anand2022}.

Satellites that have already had their CGM removed have lost their barrier or protection against the ICM. If there is continued star formation or black hole accretion, then the ensuing outflows would interact with the ICM, heating up and becoming unbound from the satellite, that is being directly deposited into and mixed with the ICM \citep{McGee2014}. As a majority of satellites in TNG-Cluster are gas-poor today, many satellites have deposited their CGM and ISM into cluster atmospheres, and these satellites could potentially act as a source of metals in cluster outskirts \citep[see][]{Nelson2024}.

Satellites that still retain their CGM behave differently. First, the CGM will roughly co-move with the satellite, shielding it from ram pressure stripping. To an extent, satellites with a spatially extended, roughly spherical CGM may continue evolving similarly to central galaxies of the same mass, that is forming stars, growing their central SMBHs, and even accreting cold gas from their own CGM. However, there remain key differences between these massive satellites and their central counterparts \citep[e.g.,][]{Engler2020}. For example, in TNG100 massive satellites $\mstarsat\sim10^{10.6-11.6}\, \msun$ that quench as satellites tend to have reduced SMBH accretion rates compared to centrals \citep{Joshi2020}. This could be caused by lower gas accretion rates from a disturbed or partially stripped CGM. The existence of the CGM for satellite galaxies modulates not only their secular evolution but also the impact of environment, as well as the strength and nature of interactions with the parent cluster ICM.

\section{Summary and main conclusions} \label{sec:sum}

In this work we study the gas content, circumgalactic medium retention, and observability of $\approx90,000$ satellite galaxies in and around 352 host clusters from the new TNG-Cluster simulation \citep[][\textcolor{blue}{Pillepich et al. in prep.}]{Nelson2024}. This simulation provides statistical and representative samples of satellites at high resolution $m_{\rm bar}\sim10^7\, \msun$ with the well-validated TNG galaxy formation model \citep{Weinberger2017,Pillepich2018b}. We focus on satellites with stellar mass $\mstarsat\sim10^{9-12.5}\,\msun$ around clusters with total mass $\mvirhost\sim10^{14.3-15.4}\,\msun$ at $z=0$. Our main results are:
\begin{itemize}
    \item The number of satellites per cluster above a given stellar mass threshold increases with cluster mass such that a cluster of mass $\mvirhost\sim10^{14.5}\, (10^{15})\, \msun$ hosts $\sim100\, (300)$ satellites today (Fig.~\ref{fig:halorichness}, \S~\ref{sec:results_demographics}). Of these satellites, $\sim 40\, (100)$ are massive $\mstarsat\sim10^{10-12.5}\msun$. Each cluster hosts at least a few extremely massive satellites $\mstarsat\sim10^{11-12.5}\, \msun$, that is,as or more massive than Andromeda. The TNG-Cluster satellite-richness relation broadly agrees with SDSS observations.
    \item Across all studied stellar and host masses, only a minority (10~percent) of satellites retain significant gas reservoirs $\mgassat>10^{9}\, \msun$ at $z=0$. The fraction of gas-rich satellites increases with satellite stellar mass (Fig.~\ref{fig:gasretention}, \S~\ref{sec:results_gascontent}), where more massive satellites have higher gas masses. There is little trend with host cluster mass.
    \item lower-mass satellites $\mstarsat\sim10^{9-10}\, \msun$ are more likely to retain, if at all, a mostly cold interstellar medium (ISM) as opposed to a hot CGM, as ram pressure preferentially removes the CGM first (Figs.~\ref{fig:satelliteCGM},~\ref{fig:SGRP_narrowbins_multipanel}, \S~\ref{sec:results_gascontent},\S~\ref{sec:results_diversity}). higher-mass satellites $\mstarsat\sim10^{10.75-12.5}\msun$ are more likely to retain a mostly hot, spatially extended CGM because of their stronger self-gravity, and the ejective impact of AGN feedback on ISM gas (Figs.~\ref{fig:satelliteCGM},~\ref{fig:SGRP_narrowbins_multipanel}). 
    \item With a sample of over 5,000 satellites that retain a sizeable amount of their CGM gas, we find that CGM gas mass increases with satellite stellar mass (Fig.~\ref{fig:satelliteCGM}). More massive satellites lose less CGM mass since infall, reflecting their resistance to environmentally driven gas-removal processes. 
    \item We predict that many gas-rich TNG-Cluster satellites should be visible in the soft X-ray (0.5-2.0~keV), even without a background subtraction (Fig.~\ref{fig:xray_poster}, \S~\ref{sec:discussion_xray}).
    \item We quantify the soft X-ray excess around satellites by subtracting a smoothed model from the total X-ray surface brightness maps (Fig.~\ref{fig:soft-xray-excess-multipanel}). When mean-stacking all $37,000$ satellites of mass $\mstarsat\sim10^{10-12.5}\,\msun$ within a projected distance $<1.5\rvirhost$, there is an excess in the X-ray surface brightness up to $\approx10\times$ the background, extending to $\approx50-100$~kpc (Fig.~\ref{fig:Lxsoft_stacks}). The excess is the largest for satellites with high masses and large cluster-centric distances.
    \item We contrast the soft X-ray contribution from satellites to the total ICM emission (Figs.~\ref{fig:soft-xray_onlySatellites_mosaic}, \ref{fig:coveringfraction}; \S~\ref{sec:discussion_absorption}). Satellites can contribute significantly to the X-ray emission over large portions of the projected areas of clusters, that is,they have a high covering fraction. Approximately $10$~percent of the soft X-ray emission in cluster outskirts $\approx0.75-1.5\rvir$ originates from satellite galaxies.
\end{itemize}

In conclusion, massive satellite galaxies are able to retain at least some of their hot, spatially extended, X-ray emitting CGM, despite living in harsh cluster environments. The gaseous atmospheres around some of these satellites should be visible in the soft X-ray, sometimes without a background subtraction. These results have numerous implications related to the evolution of satellite galaxies and interpretation of X-ray surveys and background quasar absorption studies. The presence or not of a satellite CGM affects how the ICM can remove the ISM of the satellite, how the outflows from the satellite interact with the surrounding medium, and to what extent the satellite is able to accrete from the surrounding medium. We predict a fraction of satellites do in fact retain a sizeable amount of their CGM, and these extended gas reservoirs contribute to the X-ray flux of their host clusters. The average cluster has a high covering fraction of satellite CGM, especially in the outskirts, and these CGM contribute absorbers along the line of sight to background quasars. Future studies can model the TNG-Cluster satellite CGM to compute quantitatively the column densities of specific ions, also in the UV, such as Mg\,{\sc ii} or O\,{\sc vi}, whose absorbing strength depend on the thermodynamic properties of the gas. As we expect these properties to differ between the gas in the satellite CGM and that in the ICM, the absorption features could be dominated by the interloping satellite CGM rather than the ICM. Future observational studies of both quasar absorption features and X-ray clusters should consider the possible influences from the satellite population.

\section*{Acknowledgements and Data Availability}

We thank the anonymous referee for the detailed suggestions that improved the quality of the manuscript. 

ER is a fellow of the International Max Planck Research School for Astronomy and Cosmic Physics at the University of Heidelberg (IMPRS-HD). ER would like to acknowledge the following (in alphabetical order) for useful comments and discussions that improved the quality of the manuscript: Chris Byrohl, Anna de Graaff, Morgan Fouesneau, Maximilian H{\"a}berle, Noa Hoffmann, Katrin Lehle, Joseph S. W. Lewis, Iva Momcheva, M. Selina Nitschai, Hans-Walter Rix, Nhut Truong, Nico Winkel, and Zhang-Liang Xie.

DN and MA acknowledge funding from the Deutsche Forschungsgemeinschaft (DFG) through an Emmy Noether Research Group (grant number NE 2441/1-1). Moreover, this work is co-funded by the European Union (ERC, COSMIC-KEY, 101087822, PI: Pillepich). Views and opinions expressed are however those of the author(s) only and do not necessarily reflect those of the European Union or the European Research Council. Neither the European Union nor the granting authority can be held responsible for them.

The TNG-Cluster simulation suite has been executed on several machines: with compute time awarded under the TNG-Cluster project on the HoreKa supercomputer, funded by the Ministry of Science, Research and the Arts Baden-Württemberg and by the Federal Ministry of Education and Research; the bwForCluster Helix supercomputer, supported by the state of Baden-Württemberg through bwHPC and the German Research Foundation (DFG) through grant INST 35/1597-1 FUGG; the Vera cluster of the Max Planck Institute for Astronomy (MPIA), as well as the Cobra and Raven clusters, all three operated by the Max Planck Computational Data Facility (MPCDF); and the BinAC cluster, supported by the High Performance and Cloud Computing Group at the Zentrum für Datenverarbeitung of the University of Tübingen, the state of Baden-Württemberg through bwHPC and the German Research Foundation (DFG) through grant no INST 37/935-1 FUGG. 

The IllustrisTNG simulations themselves are publicly available and accessible at \url{www.tng-project.org/data}, as described in \citet{Nelson2019}, where the TNG-Cluster simulation will also be made public in 2024. Data directly related to this publication is available on request from the corresponding author. All codes used to analyze the TNG-Cluster data and to produce the figures in this manuscript are publicly available at \url{https://github.com/ecrohr/TNG\_RPS}. 

Software used: {\sc Python} \citep{VanDerWalt2011}; {\sc IPython} \citep{Perez2007}; {\sc Numpy} \citep{VanDerWalt2011,Harris2020}; {\sc Scipy} \citep{Virtanen2020}; {\sc Matplotlib} \citep{Hunter2007}; {\sc Jupyter} \citep{Kluyver2016}. This work made extensive use of the NASA Astrophysics Data System and \url{arXiv.org} preprint server.

\bibliographystyle{aa}
\bibliography{references}

\begin{appendix}

\section{Defining a gas mass threshold for gas-rich satellites} \label{app:resolution}

\begin{figure}
    \includegraphics[width=\columnwidth]{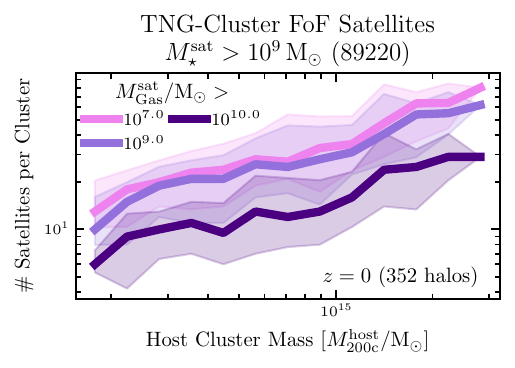}
    \caption{Similar to Fig.~\ref{fig:halorichness} (bottom left panel) except we show the number of satellites per host with a given gas mass $\mgassat$ or higher. The medians and 16th and 84th percentiles are the solid curves and shaded regions respectively. The number of satellites per host above a given gas mass increases with host mass. While $\approx90$~percent of satellites have gas masses $\mgassat < 10^{7}\, \msun$, 90~percent of those with gas mass $\mgassat \geq 10^{7}\, \msun$ ($\sim$ 1 gas cell) actually have $\mgassat \gtrsim 10^9\, \msun$ ($\approx 100$ gas cells), suggesting that the gaseous TNG-Cluster satellites are sufficiently resolved.}
    \label{fig:resolution}
\end{figure}

Fig.~\ref{fig:resolution} shows the number of satellites per host, as a function of host halo mass, with a gas mass above a given threshold. Halos with mass $\mvirhost \approx 10^{14.5}\, \msun$ host $\approx 100$ satellites (Fig.~\ref{fig:halorichness} bottom left panel), but only $\approx 10$ ($\approx10$~percent) of those galaxies retain gas masses above our resolution limit at $z=0$. In fact, the majority of satellites have negligible gas masses $\mgassat \lesssim 10^7\, \msun$. The curves for the numbers of satellites with $\mgassat > 10^{7}\, \msun$ (light purple) and $\mgassat > 10^{9}\, \msun$ (purple) are remarkably similar: there are only $\approx 10$~percent more of the former. This suggests that satellites with nonzero gas content $\mgassat > 10^{7}\, \msun$ (that is,at least one gas cell, or above the TNG-Cluster resolution limit) are reasonably resolved and relatively gas rich $\mgassat \gtrsim 10^{9}\, \msun$ ($\gtrsim100$ gas cells).

As a result, we consider satellites with $\mgassat > 10^{9}\, \msun$ gas-rich and those with $\mgassat < 10^9\,\msun$ gas-poor. We check that adding an additional criterion that the gas fraction $f_{\rm gas} \equiv \mgassat / (\mgassat + \mstarsat) > 10^{-2},\ 10^{-1}$ removes 40 (0.4~percent), 777 (7.7)~percent of the gaseous satellites, and excising these galaxies does not significantly affect the results.

Moreover, the difference between these two $\mgassat > 10^7,\ 10^9\, \msun$ (light purple, purple) populations is much smaller than between the $\mgassat > 10^9,\ 10^{10}\, \msun$ (purple, dark purple) populations. This well-defined difference suggests that the processes removing gas from satellites are spatially and temporarily resolved at gas masses $\mgas\sim10^{9-10}\,\msun$, but perhaps not at lower gas masses $\mgassat \lesssim 10^{9}\, \msun$. That is, once a gas reservoir drops to $\sim$~a~few~$\times10^2$ cells, it quickly becomes gas-poor. A resolution convergence study with the IllustrisTNG boxes can quantitatively assess possible numerical effects on satellite gas contents and physical properties \citep[e.g.,][]{Joshi2020,Donnari2021b,Stevens2021}.

\end{appendix}

\end{document}